\documentclass[10pt, a4paper, aps, prc, twocolumn, floatfix, superscriptaddress, nofootinbib]{revtex4-2}
\usepackage{graphicx}
\usepackage{float}
\usepackage{listings}
\usepackage{amsmath}
\usepackage{color}
\usepackage{bm} 
\usepackage{kantlipsum} 
\usepackage{subfiles} 
\usepackage{physics} 
\usepackage{enumitem} 
\usepackage{multirow}
\usepackage{subfig} 
\usepackage{amsfonts}
\usepackage{siunitx}
\usepackage{subcaption}
\usepackage{booktabs}
\usepackage{isotope}
\usepackage{mathtools} 

\usepackage{hyperref}
\hypersetup{
    colorlinks=true,
    linkcolor=blue,
    filecolor=magenta,      
    urlcolor=blue,
    citecolor=blue,
}
\usepackage{cleveref}

\begin{document}

\title{Microscopic study of the low-energy enhancement in the gamma-decay strength of \(^{50}\)V }
\title{Microscopic study of the low-energy enhancement in the gamma-decay strength of \texorpdfstring{$^{50}$V}{50V}}
\author{J.~K.~Dahl}
\email{j.k.dahl@fys.uio.no}
\author{A.~C.~Larsen}
\email{a.c.larsen@fys.uio.no}
\affiliation{Department of Physics, University of Oslo, N-0316 Oslo, Norway}

\author{N.~Shimizu}
\affiliation{Center for Computational Sciences, University of Tsukuba, Tsukuba, Ibaraki 305-8577, Japan}
\affiliation{Center for Nuclear Study, the University of Tokyo, Hongo, Tokyo, 113-0033, Japan}
\email{shimizu@nucl.ph.tsukuba.ac.jp}

\author{Y.~Utsuno}
\affiliation{Center for Nuclear Study, the University of Tokyo, Hongo, Tokyo, 113-0033, Japan}
\affiliation{Advanced Science Research Center, Japan Atomic Energy Agency, Tokai, Ibaraki 319-1195, Japan}
\email{utsuno.yutaka@jaea.go.jp}
\date{\today}

\begin{abstract}
    We address the microscopic origin of the low–energy enhancement (LEE) in \isotope[50]{V} with large–scale shell–model calculations to obtain $E1$ and $M1$ transitions within the same theoretical framework. The valence space spans the three major shells $sd$, $pf$ and $sdg$ and is treated with the \texttt{SDPFSDG-MU} interaction using the KSHELL code. With a \(1 \hbar \omega\) truncation, 3600 energy eigenstates and a basis of $7.02\times10^{6}$ positive and $5.94\times10^{8}$ negative parity states, the calculations yield nearly two million individual dipole transitions. The fourteen lowest experimental levels are reproduced within $0.30$~MeV, the calculated total level density excellently reproduces Oslo–method data up to $E\!\approx\!7.5$~MeV, and the calculated dipole gamma strength function follows the experimental shape — including the LEE — for the full gamma-energy range covered by the Oslo experiment. The LEE is shown to be entirely magnetic dipole in origin. Both spin and orbital parts of the \(\hat{M}1\) operator are required to reproduce the LEE, with constructive interference between the spin and orbital parts giving an extra enhancement to the LEE. Reduced one–body transition densities identify $0f_{7/2}\!\rightarrow\!0f_{7/2}$ proton transitions as the principal driver of the LEE.
\end{abstract}

\maketitle

\section{\textbf{Introduction}}
    Atomic nuclei are governed by the laws of quantum mechanics and exhibit complex behaviour, particularly at high excitation energies entering the quasi-continuum region. In this energy region, where the quantum levels are closely spaced but still not overlapping, nuclei are best described through average properties such as the nuclear level density (NLD) and the gamma strength function (GSF). The latter quantity describes the average reduced gamma-decay probability of excited nuclei, and sheds light on nuclear structure and dynamics in the quasi-continuum.
    
    At transition energies above $E_\gamma \approx 10-15$ MeV, the GSF is dominated by the giant electric dipole resonance (GDR), which, in the macroscopic picture, is due to a collective excitation mode where neutrons move against protons. However, on the low-energy tail of the GDR, other features are observed, such as the \(E1\) pygmy dipole resonance \cite{pygmy1, pygmy2}, \(M1\) scissors mode \cite{scissors}, and the low-energy enhancement (LEE) \cite{tavukcu_phd, PhysRevLett.93.142504, PhysRevLett.108.162503}. Regarding the LEE, experimental data over the last 20 years or so have revealed an unexpected increase in the GSF for gamma energies below \( \sim 4\) MeV, found in various mass regions, e.g. in Fe \cite{PhysRevLett.93.142504,PhysRevLett.111.242504,Larsen_2017,PhysRevC.97.024327}, V \cite{PhysRevC.73.064301}, La~\cite{PhysRevC.95.045805} and even rare-earth isotopes \cite{PhysRevC.76.044303}. This feature was confirmed with an independent experimental technique for the cases of $^{95}$Mo~\cite{PhysRevLett.108.162503}, $^{60}$Ni~\cite{PhysRevC.81.024319} and $^{56}$Fe~\cite{PhysRevC.97.024327}. Still, there are conflicting experimental results, specifically from two-step gamma cascades in the case of $^{96}$Mo~\cite{PhysRevC.77.054319} and from the analysis of primary $\gamma$ transitions following neutron capture in $^{57}$Fe~\cite{PhysRevC.111.044606}, although multi-cascade spectra from average resonance capture seem to be fully consistent with a magnetic-dipole type LEE~\cite{PhysRevC.99.044308}. Moreover, GSF data obtained with the beta-Oslo method, i.e. NLDs and GSFs extracted from $\gamma$ cascades following $\beta$ decay~\cite{PhysRevLett.113.232502}, have revealed the LEE also in neutron-rich nuclei such as $^{70}$Ni~\cite{PhysRevLett.116.242502,PhysRevC.97.054329} and $^{60}$Fe~\cite{SpyrouNatureComm2024}. The LEE is also known by various other terms in the literature, such as \textit{soft pole} \cite{PhysRevC.71.044307}, \textit{upbend} \cite{PhysRevC.93.034303}, and \textit{low-energy magnetic radiation} \cite{PhysRevLett.118.092502}.
    
    The underlying mechanism of the LEE is still under debate. On the theoretical side, multiple shell model studies indicate that \(M1\) transitions are responsible for the LEE. Schwengner \textit{et al.}~\cite{PhysRevLett.111.232504} demonstrated that their \(M1\) shell model GSF could explain the LEE in \(^{90}\)Zr and \(^{94, 95, 96}\)Mo, but they only included \(M1\) transitions in their calculations, estimating the \(E1\) GSF component through the Generalized Lorentzian expression. 
    A similar approach was taken for the $^{70}$Ni case, however, here the \(E1\) part was estimated within the quasiparticle time-blocking approximation (see Ref.~\cite{PhysRevC.97.054329}). On the other hand, K. Sieja used the shell model to calculate both \(E1\) and \(M1\) GSFs for \(^{44}\)Sc~\cite{PhysRevLett.119.052502}, and revealed that \(M1\) transitions were the primary contributors to the observed LEE, while \(E1\) transitions had minimal impact. Liddick \textit{et al.} also employed KSHELL to compute both \(E1\) and \(M1\) transitions in their examination of \(^{51}\)Ti's LEE comparing to both Oslo data from the $^{50}$Ti(d,p$\gamma$)$^{51}$Ti reaction and data from $^{51}$Sc undergoing $\beta^-$ decay \cite{PhysRevC.100.024624}. They reached the same conclusion that the LEE is strongly dominated by $M1$ transitions. 
    
    To date, few studies have included both \(E1\) and \(M1\) transitions in the same computational framework due to the very challenging calculations needed for a model space supporting \(E1\) transitions. The novelty of this work is to use very large shell model calculations, pushing both software and hardware to the limit, while using a model space which supports both \(E1\) and \(M1\) transitions. Our main aim is to pinpoint the electromagnetic nature of the LEE, as well as investigate the single-particle orbitals involved, for the case  of $^{50}$V. This nucleus is particularly interesting as it is the first odd-odd nucleus for which the LEE was observed, thus having a relatively high level density,  corresponding to an expected statistical $\gamma$-decay above the discrete region. Moreover, $^{50}$V is one of the below-iron-group nuclei included in reaction network calculations for e.g. core-collapse supernova nucleosynthesis~\cite{Imasheva_2023} and nucleosynthesis of massive stars~\cite{Rauscher_2002}. As $^{49}$V is unstable, there is no experimental data on the radiative neutron-capture cross section and the reaction networks rely fully on theoretical rates, which can vary by an order of magnitude or more~\cite{PhysRevLett.116.242502}. By providing a deeper understanding of both NLDs and GSFs in this mass region, these rates could be significantly improved and potentially lead to more precise nucleosynthesis calculations.
    
    The article is organized as follows. In ~\cref{sec:KSHELL_details}, we provide the details of the shell-model calculations, while in \cref{seq:nucprop} we formally introduce the nuclear properties we are interested in, namely the NLD and GSF. Further, we detail the concept of one-body transition density in \cref{seq:obtd}, before comparing the results from our calculations to experimental data in \cref{sec:quality}. In \cref{seq:GSF_features_and_mechanisms}, we present a thorough discussion of our results concerning the nature of the LEE within the shell-model calculations, in particular the role of the various single-particle orbitals that are included in our model space. Finally, we give a summary and outlook in \cref{seq:summary}.
\section{Details of the shell-model calculations}
\label{sec:KSHELL_details}
    \begin{figure}[t]
        \centering
        \hspace*{-1.7cm}
        \includegraphics[width=1.3\columnwidth]{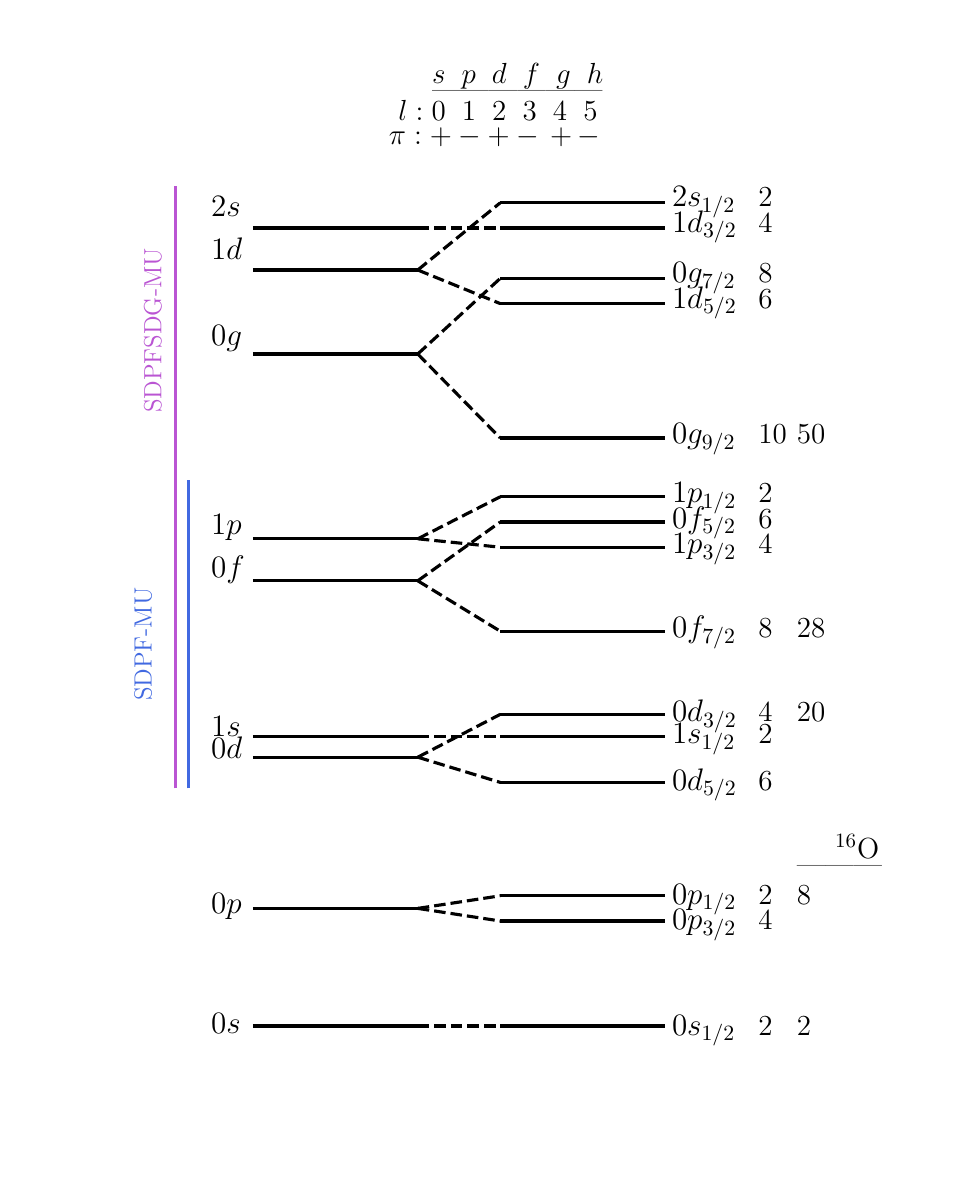}
        \caption{Approximate energy spacings of the orbitals in the nuclear shell model. The model spaces of \texttt{SDPF-MU} and \texttt{SDPFSDG-MU} are indicated in blue and purple respectively.}
        \label{fig:nuclear_shell_model}
    \end{figure}
    \subsection{Model space and effective interaction}
        For the calculations we have employed the \(M\)-scheme shell model code KSHELL \cite{KSHELL}. For each of the total angular momenta \(J = 0, 1, \ldots, 8\) and for both parities, we have calculated 200 levels totalling 3600 levels, for which there are a total of 960000 \(E1\) transitions and 958400 \(M1\) transitions. In practice we had to exclude some of the transitions due to numerical reasons\footnote{In some cases, two different levels end up having the exact same energy – to the floating point precision corresponding to an energy of 1 keV – leading to a gamma energy of 0. We have decided to exclude these relatively few transitions.}, resulting in a total of 959834 \(E1\) transitions and 938347 \(M1\) transitions. Having a large number of transitions is crucial for extracting the GSF since it describes \textit{average} transition probabilities. 

        We have taken the full $sd$-$pf$-$sdg$ shell as the valence shell, see \cref{fig:nuclear_shell_model} for an overview of the model space and orbitals. 
        In this model space, \(^{50}\)V have 15 valence protons and 19 valence neutrons. This large valence space allows us to construct the full set of \(0 \hbar \omega\) and \(1 \hbar \omega\) basis states which are needed to describe the natural and unnatural parity states respectively (see \cref{sec:truncation} for more details). It should be stressed that taking the full \(1 \hbar \omega\) model space is highly desirable to obtain reliable \(E1\) strengths because the $E1$ operator produces $1\hbar\omega$ states when applied to $0\hbar\omega$ states. It is also noted that the level densities of unnatural-parity states are close to those of natural-parity states for $E_x \gtrsim 3$ MeV as shown later, thus playing an essential role also in the low-energy $M1$ transitions.
        
        We have used the \texttt{SDPFSDG-MU} effective interaction \cite{PhysRevC.97.054321}. This interaction is based on the \texttt{USD} interaction \cite{usd} for the $sd$-shell part, the \texttt{GXPF1B} interaction \cite{gxpf1b} for the $pf$-shell part, and a variant of the $V_{\textrm{MU}}$ interaction \cite{vmu} that is employed in the \texttt{SDPF-MU} interaction \cite{sdpfmu} for the remaining part. In addition, a few two-body matrix elements are modified in the same way as the \texttt{SDPF-MU} interaction, see Ref. \cite{sdpfmu} for more details. 
        Since the Fermi surface of $^{50}$V is located in the $pf$-shell, the $pf$-shell interaction is particularly important for describing this nucleus. The \texttt{GXPF1B} interaction was constructed in \cite{gxpf1b} to achieve a good description of neutron-rich Ca isotopes with minimal modifications to the semi-empirical \texttt{GXPF1} interaction \cite{gxpf1}.

        The \(E1\) and  \(M1\) contributions are of particular interest when studying the GSF in the quasi-continuum, because dipole transitions are known to be dominant in this excitation-energy region~\cite{PhysRevC.41.1941}. Hence, it is advantageous to be able to calculate both \(E1\) and  \(M1\) GSFs within the same model framework to get a consistent theoretical description of the dipole strength, and to estimate their relative contributions to the total dipole GSF.
        
    \subsection{Truncation} \label{sec:truncation}
        The main computational challenge of a shell model calculation is that the number of basis states can become very large. To make calculations manageable, we have applied a \(1\hbar \omega\) truncation. This means that at any given time, only a single nucleon can be excited across one major shell gap. The term “\(1\hbar \omega\)” refers to a change of one unit in the harmonic oscillator quantum number, labelled \(N = 2n + l\), when a nucleon is excited or de-excited to another major shell. In the lowest energy configuration, valence nucleons occupy orbitals in order of increasing energy. This configuration has the minimum possible harmonic oscillator quantum number. Any rearrangement of nucleons within a single major shell – where they swap places without crossing a major shell gap – does not change this quantum number and is considered a valid basis state. Moving a nucleon from the \(sd\) to the \(pf\) major shell, or from the \(pf\) to the \(sdg\) shell, changes the harmonic oscillator quantum number by 1, which aligns with the \(1\hbar \omega\) truncation and is permitted as a valid basis state. In contrast, moving multiple nucleons or moving one nucleon across two major shell gaps would exceed the \(1\hbar \omega\) limit, and these configurations are therefore excluded. There are no further truncations on specific orbitals in the model space, meaning that we allow the \textit{full} set of \(0 \hbar \omega\) and \(1 \hbar \omega\) basis states.
        
        The \(1 \hbar \omega\) truncation was essential for reducing the \(M\)-scheme dimension \(D_M\) to a manageable size. Specifically, for the \(^{50}\)V calculations of this work, the \(M = 0\) dimension was \(7.02 \times 10^6\) for positive states and \(5.94 \times 10^8\) for negative states. The negative-parity dimension is still quite large despite the truncation, requiring that the calculations are parallelised over a few thousand CPU cores to keep a realistic calculation time.

    \subsection{Centre-of-mass contamination and spurious states}
        While at least two major shells are needed to calculate the \(E1\) GSF, we have used an interaction with three. Adding the \(sdg\) shell to the \(sd\) and \(pf\) shells plays an important role at removing centre-of-mass (CM) contamination. CM contamination means that physical shell-model states contain an admixture of CM-excited components \cite{GLOECKNER1974313}. The true Hamiltonian of the system is invariant under translation, however, the Hamiltonian in the truncated basis is not completely translationally invariant.  This gives rise to a CM-energy contribution to the energies of the real physical states. By using \texttt{SDPFSDG-MU} we ensure that the CM energy contribution is kept to a minimum.
        
        CM contamination in the truncated-basis Hamiltonian also produces completely non-physical states called \textit{spurious states}. As they have no physical significance, and only pollute the end-results, we want them completely removed from the excitation-energy region of interest. KSHELL handles the spurious states by use of the Lawson method~\cite{lawson1980theory} in which a shifted CM Hamiltonian, $H_{\mathrm{CM}}$, is added to the shell-model Hamiltonian:
        \begin{align}
            H = H_{\mathrm{SM}} + \beta_{\mathrm{CM}}\left(H_{\mathrm{CM}} - \frac{3}{2} \hbar \omega\right).
        \end{align}
        By using an adequately large value \(\beta_{\mathrm{CM}}\), the spurious states are pushed to much higher energies, effectively removing them from the energy region of interest, while only adding a negligible CM-energy contribution to the real physical states. Specifically, KSHELL uses the same definition as the shell-model codes NushellX~\cite{BROWN2014115} and OXBASH~\cite{OXBASH} for the Lawson parameter \(\beta_{\mathrm{CM}}\). In the calculations of this work we have used
        \begin{align}
            \beta_{\mathrm{CM}} \hbar \omega / A = 10 \text{MeV},
        \end{align}
        where $A$ is the mass number of the nucleus.
        
\begin{figure}[tb]
            \centering
            \includegraphics[clip,width=1.\columnwidth]{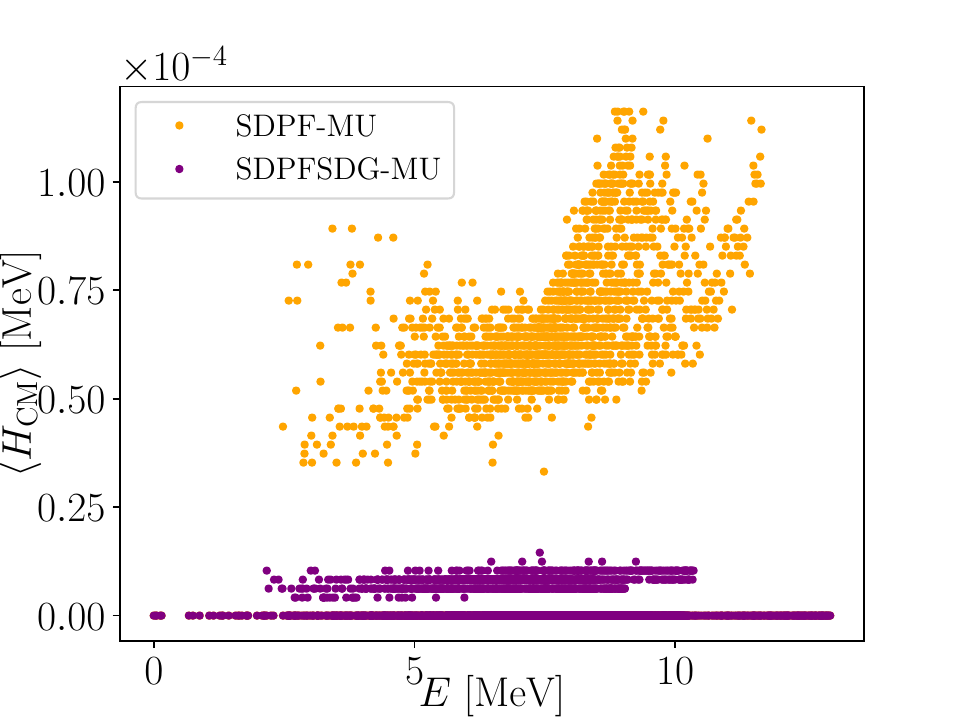}
            \caption{The expectation value of the CM Hamiltonian as a function of excitation energy for each of the 3600 levels calculated with the \texttt{SDPFSDG-MU} interaction (purple) and the \texttt{SDPF-MU} interaction (orange).}
            \label{fig:com}
        \end{figure}

        In \cref{fig:com}, we show a comparison of the CM contamination for the interactions \texttt{SDPFSDG-MU} which uses the \(sd\)-\(pf\)-\(sdg\) model space and \texttt{SDPF-MU} which uses the \(sd\)-\(pf\) model space, where the former is consistently lower than the latter. Unlike \(sd\)-\(pf\)-\(sdg\), the \(sd\)-\(pf\) model space does not cover the \(1 \hbar \omega\) basis states that has nucleons in the \(sdg\) shell, preventing one from separating the physical and spurious states. The \(sd\)-\(pf\)-\(sdg\) model space however, covers the full set of \(1 \hbar \omega\) basis states.

    \subsection{\texorpdfstring{\(g\) values for the \(M1\) operator}{g values for the M1 operator}} \label{sec:g_values}
        The \(M1\) operator is given by 
        \begin{equation} \label{eq:M1_operator}
            \hat{M}1 = \sqrt{\frac{3}{4\pi}} \mu_N(g_l \hat{L} + g_s \hat{S}), 
        \end{equation}
        where \( \hat{L} \) and \( \hat{S} \) are the orbital and spin angular momentum operators respectively, \(g_l\) and \(g_s\) are the orbital and spin \(g\) factors respectively, while \(\mu_N\) is the nuclear magneton. This operator is used to calculate the reduced transition probability \(B(M1)\), namely
        \begin{equation} \label{eq:B}
            B(M1) = \frac{|( \Psi_f \lVert \hat{M}1 \lVert \Psi_i )|^2}{2J_i + 1},
        \end{equation}
        where \(| \Psi_f \rangle\) and \(| \Psi_i \rangle\) are the wave functions for the final and initial state respectively, the initial state having total angular momentum \(J_i\). 

        The \(g_s\) values were quenched with a factor of 0.9 and we have used \(g_l = (1.1, -0.1)\) for protons and neutrons respectively following the effective \(\hat{M}1\) operator of Ref.~\cite{PhysRevC.69.034335}. The experimental ground-state magnetic moment of \(^{50}\)V is \(+3.35 \, \mu_N\) \cite{bnl_50v} while the calculated magnetic moment is \(+2.99 \, \mu_N\) showing reasonable agreement.

        The orbital and spin angular momentum parts of \cref{eq:M1_operator} can constructively and destructively interfere with each other. If the \(L\) and \(S\) terms have the \textit{same sign} they will constructively interfere with each other, and if they have \textit{different signs} they will destructively interfere. This can be seen by
        \begin{align}
            |\underbrace{( \Psi_f \lVert g_l \hat{L} \lVert \Psi_i )}_{\coloneq M_l} + \underbrace{( \Psi_f \lVert g_s \hat{S} \lVert \Psi_i )}_{\coloneq M_s}|^2 = M_l^2 + M_s^2 + 2M_lM_s
        \end{align}
        where we assume that \(M_l\) and \(M_s\) are real. The term \(2M_lM_s\) will be positive and add to the sum if the signs are equal and it will be negative and subtract from the sum if the signs are different.
        
        To understand how the \(L\) and \(S\) terms combine to form the full \(M1\) strength, it is helpful to quantify the \textit{interference angle} between the two. Analogous to the inner product of two vectors, we define
        \begin{align} \label{eq:interference-angle}
            \cos \theta = \frac{M_lM_s}{|M_l||M_s|},
        \end{align}
        where, for a single transition, the inference angle \(\theta\) is either \(0^\circ\) or \(180^\circ\) depending on whether the signs of \(M_l\) and \(M_s\) are equal or different respectively.

\section{Nuclear properties}
\label{seq:nucprop}
    The nuclear level density characterises how many nuclear energy levels lie within an excitation-energy interval \(\Delta E\). The NLD is defined as
    \begin{align}
        \rho(E) = \frac{\Delta N}{\Delta E},   
    \end{align}
    where \(\Delta N\) is the number of levels inside the bin \(\Delta E\) at excitation energy \(E\). Furthermore, each level is specified not only by \(E\) but also by its total angular momentum \(J\) and parity \(\pi\). The \textit{partial} level density, \(\rho(E, J, \pi)\), resolves the NLD by \(E\), \(J\), and \(\pi\). The total NLD is then obtained by summing over angular momentum and parity,
    \begin{align}
        \rho(E) = \sum_{J,\pi} \rho(E, J, \pi).
    \end{align}

    The gamma strength function is a statistical property of nuclei that describes average electromagnetic transition probabilities. The GSF is given by~\cite{Bartholomew1973} 
    \begin{align} \label{eq:gamma_strength_function}
         f_{\hat{\bm{O}}_\lambda}&(E_{\gamma}, E_i, J_i, \pi_i)
        \\ \nonumber
        &= \frac{\langle \Gamma^{\hat{\bm{O}}_\lambda} \rangle(E_{\gamma}, E_i, J_i, \pi_i) }{E_{\gamma}^{2\lambda + 1}}\rho(E_i, J_i, \pi_i),
    \end{align}
    where $E_i$, \(J_i\), and \(\pi_i\) are the excitation-energy, total angular momentum and parity of the initial level in a transition, while \(E_\gamma\) is the energy of the emitted gamma ray and $\hat{\bm{O}}_\lambda$ is one of the electromagnetic transition operators of rank \(\lambda\). Further, \(\Gamma\) is the partial decay width. For dipole transitions, i.e. $E1$ and $M1$ decays, the expression for the GSF becomes
     \begin{align} \label{eq:gamma_strength_M1}
        f_{\hat{\bm{O}}_1}&(E_{\gamma}, E_i, J_i, \pi_i)
        \\ \nonumber
        &= \dfrac{16 \pi}{9 \hbar^3 c^3}\langle B(\hat{\bm{O}}_1;\downarrow) \rangle (E_{\gamma}, E_i, J_i, \pi_i) \rho (E_i, J_i, \pi_i).
    \end{align}
    The GSF contains the \textit{partial} nuclear level density \(\rho\), as well as \textit{average} \(B\) values, both of which are evaluated within an excitation energy bin and gamma-ray energy bin. The \(B\) values are the \textit{reduced transition probabilities} calculated with the operator $\hat{\bm{O}}_1$, which is either the \(M1\) or \(E1\) operator for dipole transitions. The bin is taken over the excitation energy of the \textit{initial} levels of the transitions, not the final levels. Note that in some works during the years 2013 – 2017 \cite{PhysRevLett.119.052502, Brown2014, PhysRevLett.118.092502, PhysRevLett.111.232504}, the \textit{total} level density was used instead of the \textit{partial}, leading to an artificial increase in the GSF. See the appendix of \cite{PhysRevC.98.064321} for more details. In shell-model codes, the \(B\) values are commonly calculated with the help of the one-body transition density, which will be discussed in the following section.

\section{One-body transition density}  
\label{seq:obtd}
    A one-body operator in occupation number representation is given by \cite{suhonen}
    \begin{align} \label{eq:one_body_transition_operator_second_quantisation}
           \hat{O} = \sum_{\alpha \beta} \langle \alpha | \hat{O} | \beta \rangle \hat{c}^\dagger_\alpha \hat{c}_\beta.
    \end{align}
    Here, $\alpha$ and $\beta$ represent  the quantum numbers for a specific $m$ state, and $\hat{c}^\dagger_\alpha$,  $\hat{c}_\beta$ are the creation and annihilation operators, respectively. When calculating electromagnetic decay probabilities, we need to calculate the matrix elements of \cref{eq:one_body_transition_operator_second_quantisation}, namely
    \begin{align} \label{eq:matrix_element}
        \langle \Psi_f | \hat{O} | \Psi_i \rangle = \sum_{\alpha \beta} \langle \alpha | \hat{O} | \beta \rangle \langle \Psi_f | \hat{c}^\dagger_\alpha \hat{c}_\beta | \Psi_i \rangle,
    \end{align}
    where \(| \Psi_f \rangle\ = | J_f M_f \rangle \) and \(| \Psi_i \rangle = | J_i M_i \rangle \). The right inner product in \cref{eq:matrix_element} is called the \textit{one-body transition density} (OBTD). In shell-model codes it is common to use the \textit{reduced} one-body transition density (rOBTD), which we get by applying the Wigner-Eckart theorem to \cref{eq:matrix_element}, giving us 
    \begin{align} \label{eq:reduced_matrix_element}
        (J_f \lVert \hat{\bm{O}}_\lambda \lVert J_i ) = \hat{\lambda}^{-1} \sum_{a b} ( a \lVert \hat{\bm{O}}_\lambda \lVert b ) ( J_f \lVert [c_a^{\dagger} \tilde{c}_b]_\lambda \lVert J_i),
    \end{align}
    where $ \hat{\lambda} = \sqrt{2\lambda+1}$, and $a,b$ represent the quantum numbers of a specific $j$ orbital, but not a specific $m$ substate. See \cref{sec:red-trans-prob} for the definitions of the bracket and tilde operators. Further, the term $( a \lVert \hat{\bm{O}}_\lambda \lVert b )$ is the reduced single-particle matrix element, while the left-hand side in \cref{eq:reduced_matrix_element} is called the reduced matrix element, and relates to the reduced transition probability $B(\hat{\bm{O}}_\lambda)$ by
    \begin{align}
        B(\hat{\bm{O}}_\lambda) = \dfrac{|( J_f \lVert \hat{\bm{O}}_\lambda \lVert J_i )|^2}{2J_i + 1}
    \end{align}
    where \(\hat{\bm{O}}_\lambda\) is the \(E\) or \(M\) operator of multipolarity \(\lambda\). 
    The right inner product in \cref{eq:reduced_matrix_element} is the rOBTD,
    \begin{align} \label{eq:robtd}
        \rho_{fi}(a, b) = ( J_f \lVert [c_a^{\dagger} \tilde{c}_b]_\lambda \lVert J_i).
    \end{align}

    The OBTD measures how much the final state \(|\Psi_f\rangle\) overlaps with the initial state after removing a nucleon from \(m\) sub-state \(\beta\) and inserting one in \(\alpha\). A large OBTD magnitude therefore indicates that the \(\beta\!\to\!\alpha\) one-particle excitation is an important contributor to the full \(|\Psi_i\rangle\!\to\!|\Psi_f\rangle\) transition. A large overlap does however not guarantee a large observable (e.g. \(B(M1)\)), because the single-particle matrix element and phase coherence with other \(\alpha, \beta\) combinations still matter.

    The rOBTD plays the same diagnostic role as the ordinary OBTD, but with two refinements: First, coupling the creation–annihilation pair to a multipole rank \(\lambda\) and applying the Wigner–Eckart theorem removes the \(M\) dependence. 
    The rOBTD therefore tells us how strongly the orientation-independent part of the \(\beta\!\to\!\alpha\) (de-)~excitation contributes to the \(|\Psi_i\rangle\!\to\!|\Psi_f\rangle\) transition. Second, by choosing \(\lambda\) we isolate only those particle–hole components that are relevant for an operator of the same rank e.g. \(\lambda = 1\) for \(M1\), meaning that angular momentum selection rules are baked into the rOBTD. A large rOBTD, negative or positive, means that  \(b\!\to\!a\)  is a major component within that multipole channel. This is in contrast to the ordinary OBTD which does not take into account angular momentum selection rules. 
    
    By letting \(a\) and \(b\) run over all possible combinations of orbitals, we can get insight to how well the transition from \(| \Psi_i \rangle\) to \(| \Psi_f \rangle\) is described by single-particle transitions, and hence which orbitals are most responsible for this transition. By making a selection of transitions based on some characteristics, like angular momentum, parity and gamma energy, we can use the OBTD to see the involvement of single-particle transitions for that selection of transitions. For example, from a shell model calculation, we might select all the transitions which are within the gamma energy region of the LEE and which are part of the \(M1\) GSF. 
    If we then look at the OBTDs of the selected transitions we can get information about which single-particle transitions are involved in the LEE specifically.
\section{Comparison with experimental data} \label{sec:quality}
    \subsection{Discrete levels}        
\begin{figure}[!t]
            \centering
            \includegraphics[width=\linewidth]{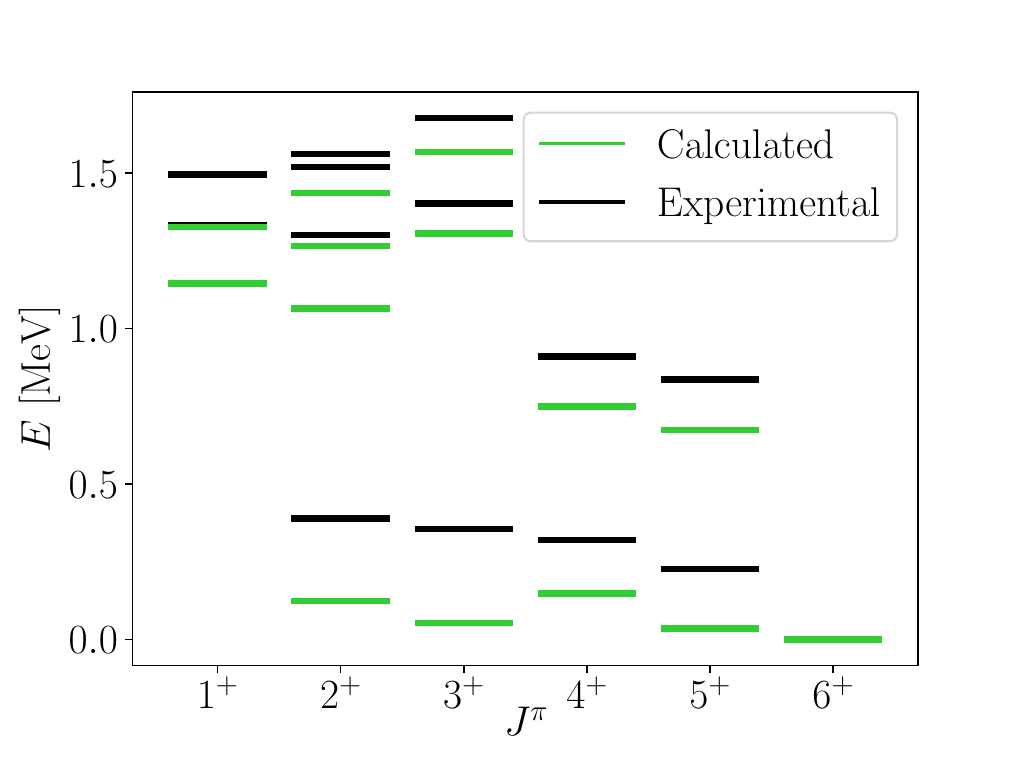}
            \caption{The 14 lowest experimental levels of \(^{50}\)V \cite{bnl_50v} (black) compared to the corresponding calculated levels (green).}
            \label{fig:level_scheme_comparison}
        \end{figure}

        A natural place to start probing the quality of a shell model calculation is to check how well it predicts known, discrete energy levels. We show such a comparison in \cref{fig:level_scheme_comparison} where a level scheme comprising the 14 lowest experimentally measured energy levels of known \(J^\pi\) is compared to the corresponding calculated energy levels. The ground state is correctly calculated to be a \( 6^+ \) level, and while the energy order of the calculated levels do not match perfectly, the \(J^\pi\)s of all the 14 lowest levels are reproduced. Between the lowest 14 calculated and experimental levels, the maximum difference is 0.30 MeV, while the mean difference is 0.19 MeV. These differences are deemed sufficiently small, since the excitation-energy bin size of \(\Delta E_i = 0.20\) MeV used in this work for the NLD and GSF is of comparable size.
        
        \begin{figure}[!t]
            \centering
            \includegraphics[width=\linewidth]{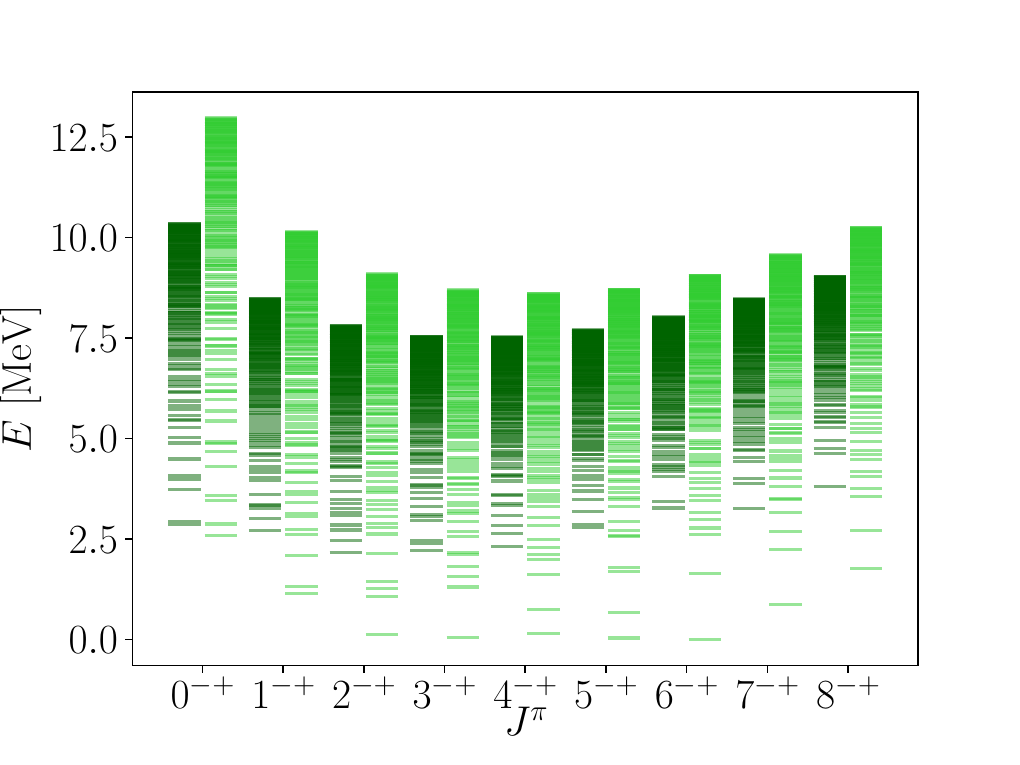}
            \caption{A level scheme of all the calculated levels of this work. The negative and positive levels are in the left and right column for each \(j\) respectively.}
            \label{fig:level_scheme}
        \end{figure}
    
        \Cref{fig:level_scheme} shows all the calculated levels; this plot allows us to see both the limitations of the calculation and the properties of the model space. The \(^{50}\)V nucleus has 23 protons and 27 neutrons, and in the single-particle picture, the non-paired neutron and proton are both in the \(0f_{7/2}\) orbital – which has negative parity – resulting in a ground state with positive parity. The next most accessible orbitals after \(0f_{7/2}\) are \(1p_{3/2}\), \(0f_{5/2}\), and \(1p_{1/2}\), all of which are in the \(pf\) major shell and all with negative parity. To make a negative parity \textit{level}, a nucleon has to either be excited from the \(sd\) shell to the \(pf\) shell, or from the \(pf\) shell to the \(sdg\) shell, both options requiring a relatively large amount of energy compared to single-particle excitations within the \(pf\) shell. It is therefore reasonable to expect that the first few excited levels are of positive parity. The lowest negative parity levels should appear at a relatively large energy, and this is indeed what we see in \cref{fig:level_scheme}. The first few calculated negative parity levels are \(2^-\), \(3^-\), and \(4^-\) at  2.16, 2.20, and 2.31 MeV, respectively. The energies coincide well with the first experimentally known negative levels, which are found at 2.16, 2.42, and 2.52 MeV and are either \(3^-\) or \(4^-\) \cite{bnl_50v}.
    
        \Cref{fig:level_scheme} reveals the upper limit of the present shell-model calculations. At approximately 7.5 MeV, there are no more \(3^-\) nor \(4^-\) levels. Other levels of energy higher than 7.5 MeV that \textit{can} decay to a \(3^-\) or a \(4^-\) level will artificially have fewer decay options, limited by the calculations and not physics, meaning that the decay probabilities of levels higher than 7.5 MeV will be incomplete. The consequences of this fact will be further discussed in the following subsections.
    
    \subsection{Nuclear level density}
        
        \begin{figure}[!t]
            \centering
            \includegraphics[width=\linewidth]{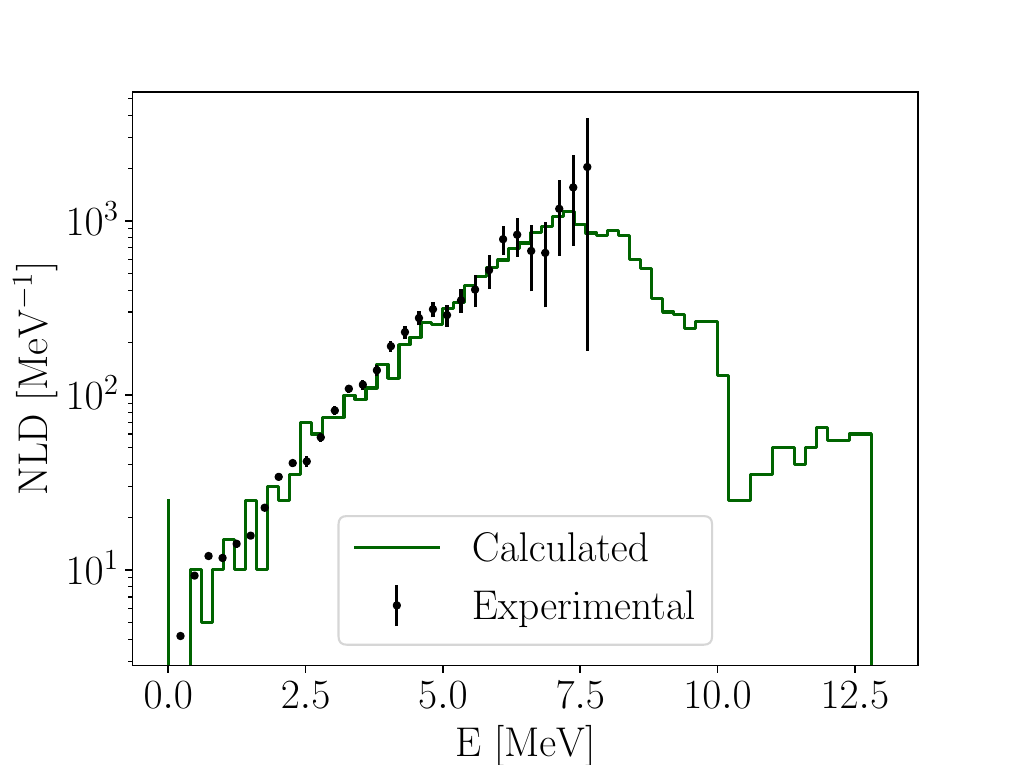}
            \caption{Comparing the experimental level density of \(^{50}\)V \cite{PhysRevC.73.064301} with the calculated level density from this work.}
            \label{fig:level_density}
        \end{figure}
        
        In \cref{fig:level_density}, we compare the total NLD from the KSHELL calculations to experimental data from the Oslo Cyclotron Laboratory~\cite{PhysRevC.73.064301} for the $^{51}$V($^3$He,$\alpha \gamma$)$^{50}$V reaction. The experimental data points cover a broad angular-momentum range and both parities. The calculations fit very well to the experimental data, all the way from excitation energy $E=0$ MeV to approximately 7.5 MeV. At 7.5 MeV, we see that the calculated NLD is no longer increasing exponentially and starts to fall off because the requested number of levels from the shell model calculation (in total 3600 levels) is no longer enough to keep up with the exponential increase. Recall as previously mentioned, that 7.5 MeV is also the point at which there are no more \(3^-\) nor \(4^-\) levels in the calculation. Note that the experimental data cuts off at approximately 7.5 MeV, due to the way the experimental NLD is extracted from the $\alpha\gamma$ coincidences.
        
        From \(E\approx 2\) MeV up to about 8 MeV the level density grows almost exponentially with excitation energy, which is exactly what the constant-temperature (CT) formula predicts~\cite{ERICSON1959}: $\rho_{CT}(E) = \frac{1}{T}\exp[(E-E_0)/T]$, where $T$ is the nuclear temperature and $E_0$ an energy shift related to the nucleon pairing. Below \(\approx 3\) MeV we enter the discrete level region and we see that the NLD fluctuates significantly. All in all, the experimental level density is very well reproduced by the shell-model calculations.

\begin{figure}[!t]
            \centering
            \includegraphics[width=\linewidth]{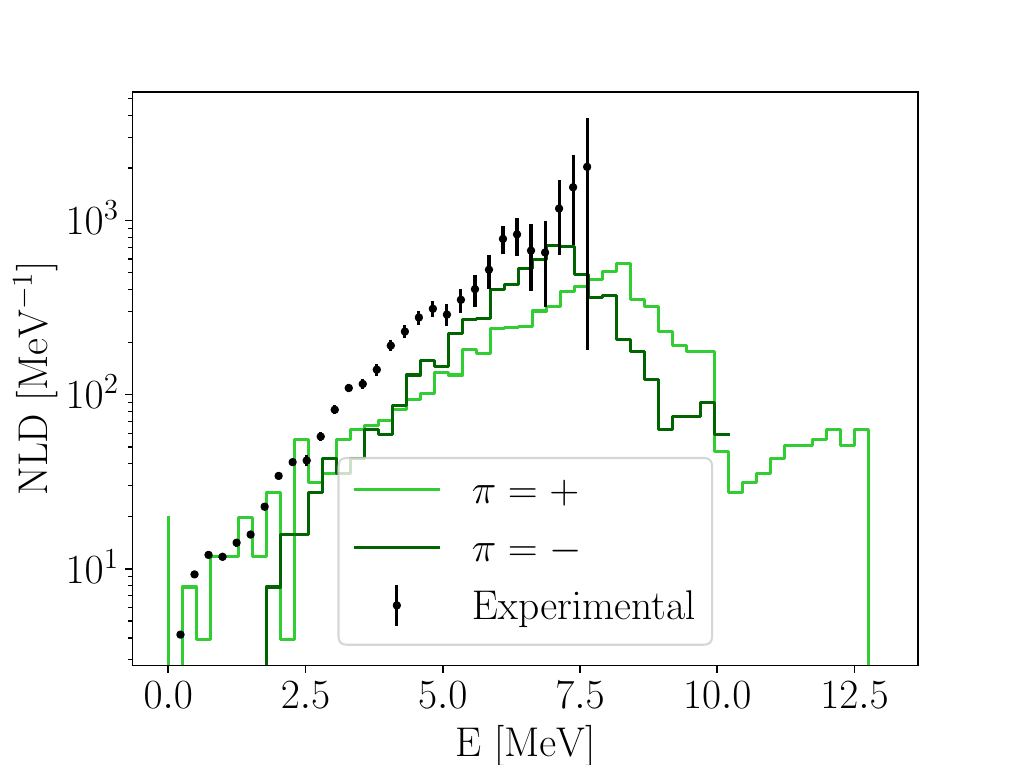}
            \caption{The calculated level density separated in parity.}
            \label{fig:level_density_separate_parity}
        \end{figure}

        From the shell-model calculations we can easily separate the calculated levels based on parity, as seen in \cref{fig:level_density_separate_parity} where we have the positive parity level density in light green and negative in dark green. Since the natural parity of \(^{50}\)V is positive, we expect that the lowest excited levels are also positive. As shown in \cref{fig:level_scheme}, the positive levels start from 0 MeV while the negative levels start at approximately 2 MeV. Negative-parity states require at least one \(\hbar \omega\) major-shell crossing, so they are energetically out of reach at low excitation energy. Once the first cross-shell possibility opens, every additional particle–hole pair needs also energy enough to account for the pairing energy. The positive-parity NLD is larger than the negative-parity NLD up to about 4 MeV from where the negative-parity NLD stays larger for the rest of the ``complete'' energy range (up to about 7.5 MeV).
        
        It is very interesting that the NLDs for the different parities have different slopes, indicating a different temperature $T$ for the two components. However, the relatively shallow slope for the positive-parity levels at approximately \(E = 5.0 – 7.5\) MeV may be due to missing \(2 \hbar \omega\) basis states in the shell-model calculation. On the other hand, as shown in \cref{fig:level_density}, our calculations agree remarkably well with the experimental data, implying that not many levels are missing from the calculations.
        
    \subsection{Gamma strength function}
        \begin{figure}[b]
            \centering
            \includegraphics[width=\linewidth]{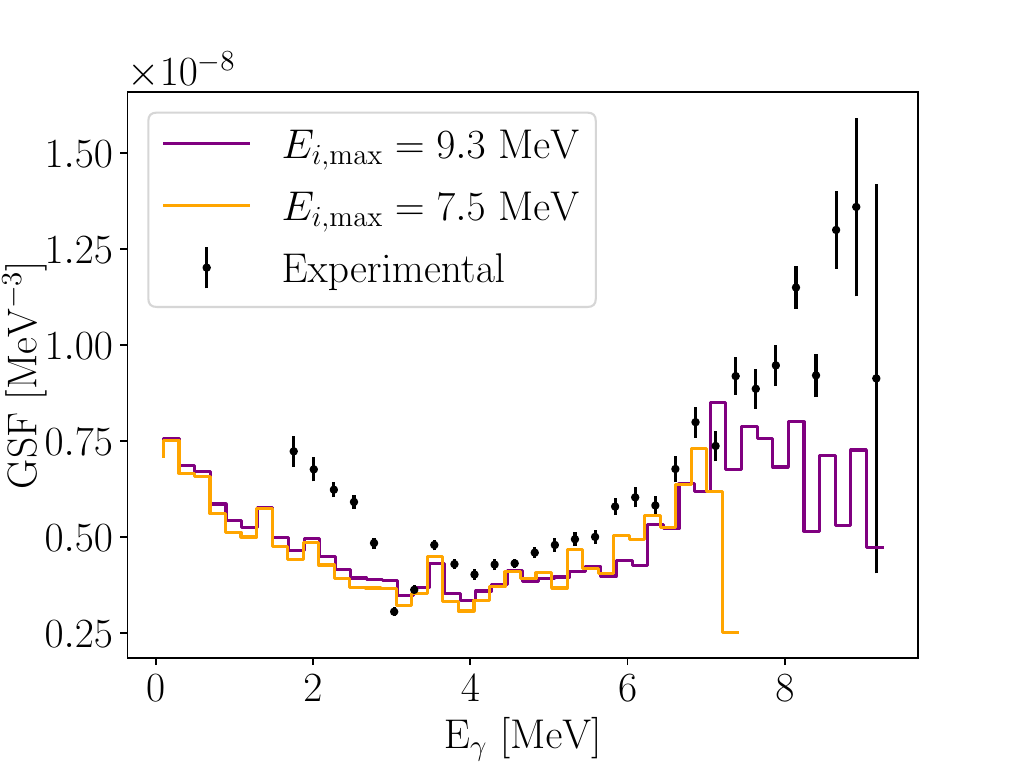}
            \caption{The calculated dipole strength function with upper excitation energy limits of \(7.5\) MeV and \(9.3\) MeV in orange and purple respectively. Experimental data in black \cite{PhysRevC.73.064301}. Note that the $y$ axis here is linear, while it is logarithmic in \cref{fig:gsf_M1_E1_normal}.}
            \label{fig:gsf_compare_75_93}
        \end{figure}
        
        \begin{figure*}[t]
            \centering
            \includegraphics[scale=0.5]{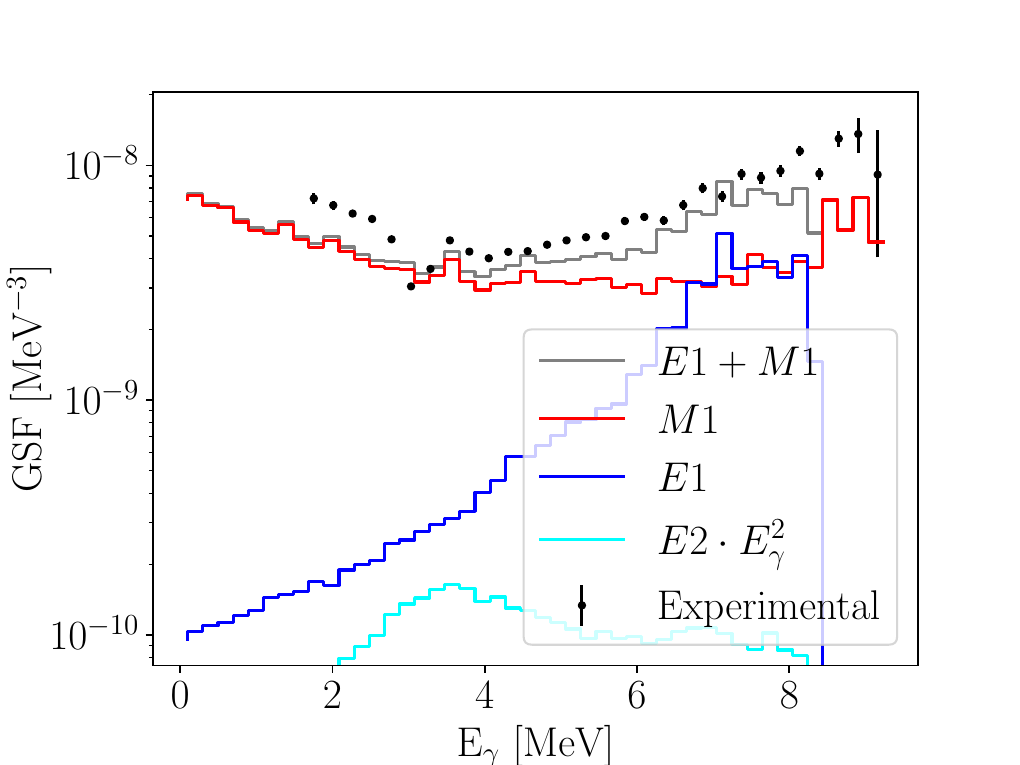}
            \includegraphics[scale=0.5]{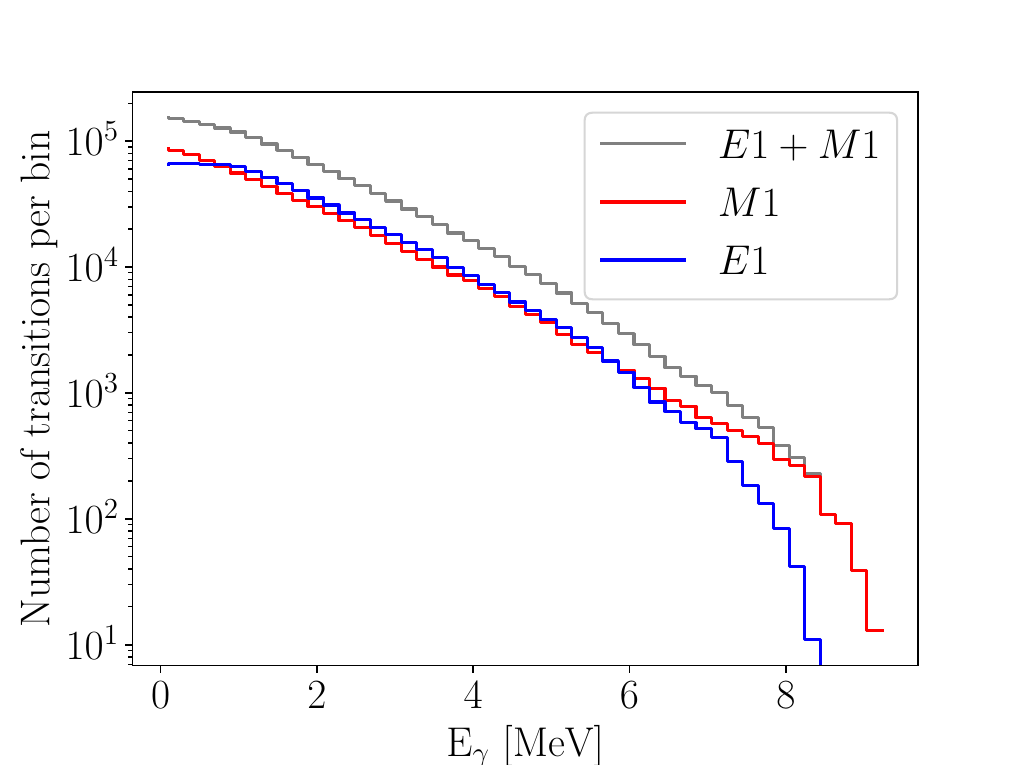}
            \caption{Left: The calculated gamma strength function of \(^{50}\)V. The total dipole strength function is seen in grey, while the individual \(E1\) and \(M1\) strength functions are seen in blue and red respectively, with experimental data in black \cite{PhysRevC.73.064301}. The \(E2\) strength function – multiplied by \(E_\gamma^2\) – is shown in cyan. Right: The number of transitions per bin used for calculating the GSF in that bin.}
            \label{fig:gsf_M1_E1_normal}
        \end{figure*}

        To perform a GSF calculation there are a few parameters that have to be chosen. In this work, the GSF calculations are limited to include transitions of initial excitation energies within a lower and an upper limit, namely \(E_{i, \text{min}} = 3.29\) MeV and \(E_{i, \text{max}} = 9.3\) MeV. A lower limit is set to make sure that the transitions included in the calculations are above the discrete-level region. The value of 3.29 MeV is chosen because it is the lower limit used in the Oslo-method experiment \cite{PhysRevC.73.064301}. The upper limit of 9.3 MeV corresponds to the neutron separation energy of \(^{50}\)V; this limit ensures that the gamma decays are not competing with neutron emission in the experimental data. An upper limit to the initial excitation energy is also an upper limit to the gamma energy since a level cannot decay with a higher energy than its  excitation energy. The upper limit also affects the entire GSF since a level at the upper limit can decay with any gamma energy from 0 to the upper limit, provided that appropriate lower levels are available. The GSF is calculated within an (initial) excitation energy bin of size \(\Delta E_i = 0.20\) MeV.

        In \cref{fig:gsf_compare_75_93}, we see the dipole strength function calculated with \(7.5\) MeV as the upper limit (orange) and with \(9.3\) MeV as the upper limit (purple). Using an upper excitation energy limit of \(7.5\) MeV is the safest choice since the exponential trend of the NLD is reproduced remarkably well up to \(7.5\) MeV, as discussed in the previous subsection. Still, it is tempting to use \(9.3\) MeV as the upper limit since a higher limit comes with a couple of advantages: Using a higher upper limit means that we can calculate the GSF to higher energies and it means that there are more transitions across the entire gamma-energy region, increasing the statistics and improving the mean. It seems that the \(E_{i, \mathrm{max}} = 9.3\) MeV strength function is slightly smoother than its counterpart because of the increased statistics. The shapes of the GSFs are more or less the same all the way up to \(7.5\) MeV, indicating that no major inconsistencies are introduced with a higher energy limit. We have therefore chosen to use \(9.3\) MeV as the upper limit from here on.
        
        We now turn  to \cref{fig:gsf_M1_E1_normal,} where we show the calculated \(M1\) (red), \(E1\) (blue), and dipole (gray) GSFs of \(^{50}\)V compared with the experimental GSF. The calculated GSFs are quite smooth up to approximately \(E_\gamma = 7.5\) MeV where we see that they start to fluctuate. The gamma-energy region where the fluctuations become significant approximately coincide with the excitation energy where the calculated NLD is no longer increasing exponentially and starts to drop off. Thus, we must be cautious about the GSF calculations at the very highest \(E_\gamma\) because the statistics  might be insufficient for producing reliable results. Recall from \cref{eq:gamma_strength_function} that the GSF describes \textit{average} electromagnetic transition probabilities; if we want the empirical mean of the GSF calculations to approach the true average, we must make sure that enough data is included in the mean. The right figure in \cref{fig:gsf_M1_E1_normal} shows the number of \textit{transitions} that were used in the GSF calculations. The number of transitions is \(\approx 10^5\) per bin for the lowest gamma energies, but steadily decreases with increasing \(E_\gamma\). At the highest \(E_\gamma\), the number of transitions drop well below $100$, which is the reason why the GSF in the same energy region has large fluctuations. When it comes to the LEE energy region, the statistics is indeed very high and we deem that the calculated GSF in the LEE energy area is reliable.

\section{\texorpdfstring{The GSF of \(^{50}\)V: features and physical mechanisms}{The GSF of 50V: features and physical mechanisms}}
\label{seq:GSF_features_and_mechanisms}
    From the left panel of \cref{fig:gsf_M1_E1_normal}, it is immediately apparent that the calculated total dipole GSF displays a \textit{low-energy enhancement} from approximately 3 MeV and down to 0 MeV. We can also clearly see that the LEE is due to the \(M1\) part of the GSF, as the \(E1\) GSF gives a negligible contribution in this energy region. While the \(E1\) GSF starts at very low values, it steadily increases as \(E_\gamma\) increases. At approximately 3 MeV there is a visible \(E1\) contribution to the total strength, and by approximately 7 MeV the \(E1\) GSF is at the same magnitude as the \(M1\) GSF. It might seem like the \(E1\) GSF decreases after 7 MeV; however, remember from \cref{sec:quality} that the GSF at the highest energies cannot be completely trusted because of low statistics and 
    low level density.

    While the amplitude of the calculated dipole strength function does not completely match the experimental data, we can see that the overall shape is successfully replicated in the calculations. We also remark that the absolute value of the experimental data has a substantial uncertainty. This is because the total average radiative width has not been measured for \(^{50}\)V \cite{PhysRevC.73.064301}. Both the calculated and experimental GSFs decrease from low to mid \(E_\gamma\), flatten out, and then increase from mid to high \(E_\gamma\). A small peak-like structure between \(3\) and \(4\) MeV is also reproduced by the shell-model calculations.

    In the Oslo method, it is usually assumed that dipole transitions are dominating and the contributions from higher multipolarities are negligible. This assumption is supported by angular-distribution measurements in this mass region~\cite{PhysRevLett.111.242504,Larsen_2017}. As shown in \cref{fig:gsf_M1_E1_normal}, the \(E2\) contribution is negligible and can safely be ignored, supporting that it is reasonable to assume dominance of dipole transitions for this nucleus in this energy region.

\begin{figure}[t]
        \centering
        \includegraphics[width=\linewidth]{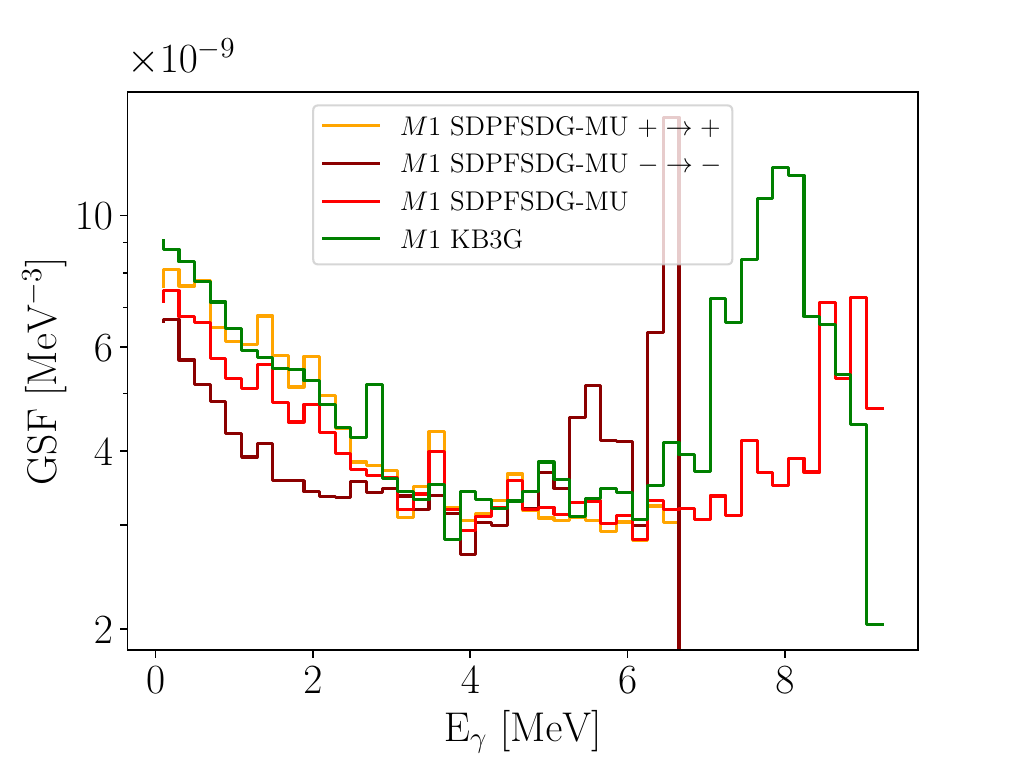}
        \caption{\(M1\) strength functions calculated with the interactions \texttt{SDPFSDG-MU} and \texttt{KB3G} in red and green, respectively. In orange and dark red we have \(M1\) strength functions with \texttt{SDPFSDG-MU} where only positive-to-positive and negative-to-negative parity transitions have been used, respectively.}
        \label{fig:gsf_compare_sdpfsdgmu_kb3g_split_parity}
    \end{figure}

    To test the impact of the chosen interaction and model space, we have also performed calculations with the \texttt{KB3G} interaction \cite{POVES2001157}. The same quenching of 0.9 as well as \(g_l = 1.1, -0.1\) was used. In \cref{fig:gsf_compare_sdpfsdgmu_kb3g_split_parity} we compare the \(M1\) strength of \texttt{KB3G} (green) and \texttt{SDPFSDG-MU} (red). In the LEE region they are generally very similar, however the \texttt{KB3G} \(M1\) strength is slightly higher. The discrepancy can be explained by the fact that \texttt{KB3G} uses the \(pf\) shell as model space and thus only captures positive-to-positive parity transitions but not negative-to-negative parity transitions. On the other hand, the \texttt{SDPFSDG-MU} interaction extends over more than one major shell and captures both types of transitions, giving us a more complete picture of the \(M1\) strength.
    
    To illustrate the implications, in \cref{fig:gsf_compare_sdpfsdgmu_kb3g_split_parity}, we have plotted separately the positive-to-positive and negative-to-negative parity \(M1\) strengths of \texttt{SDPFSDG-MU}. The positive-to-positive LEEs of the two interactions are very similar in both shape and amplitude, which shows that the calculations are quite robust with respect to the chosen interaction. Interestingly, the negative-to-negative transitions contribute to a reduction in the LEE, an effect which is not captured by a one-major-shell interaction. In the following, we will investigate in detail the nature of the LEE for the case of $^{50}$V.

    \subsection{\texorpdfstring{The role of the orbital and spin part of the \({\hat{\textbf{M}}}\textbf{1}\) operator}{The role of the orbital and spin part of the M1 operator}}
        \begin{figure}[t]
            \centering
            \includegraphics[width=\linewidth]{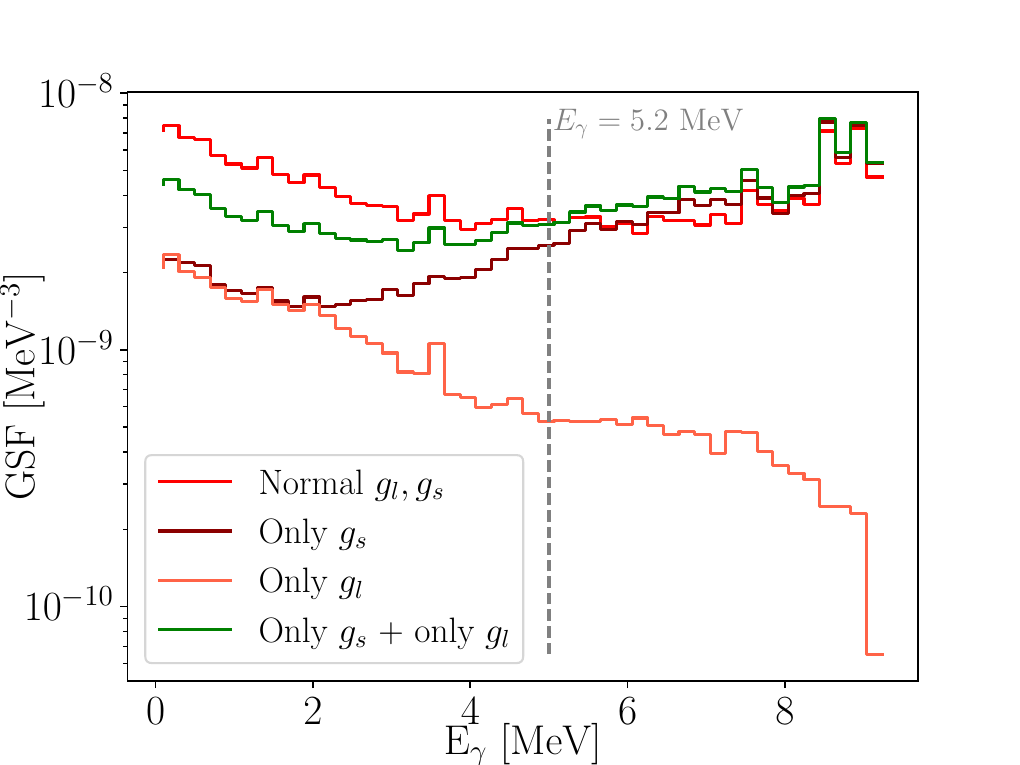}
            \caption{The \(M1\) GSF of \(^{50}\)V calculated for three sets of parameters. In red, \(g_s = 5.027, -3.443\) has been used which are the free \(g_s\) factors multiplied by a quenching factor of 0.9, while \(g_l = 1.1, -0.1\). In dark red the \(g_l\) factors are set to zero, keeping only the \(g_s\) values, while in orange the \(g_s\) factors are set to zero keeping only the \(g_l\) values. In green is the sum of ``Only \(g_s\)'' and ``Only \(g_l\)''.}
            \label{fig:gsf_M1_gl_gs_normal}
        \end{figure}

        We have seen from \cref{fig:gsf_M1_E1_normal} that the LEE is caused by magnetic dipole transitions. We can investigate the LEE further by turning off the orbital angular momentum and spin contributions to the \(\hat{M}1\) operator by setting the corresponding \(g\) factors to zero (see \cref{eq:M1_operator}). The \(\hat{M}1\) operator essentially measures how much the initial and final states are connected by the magnetic dipole operator, and by tweaking the \(g_l\) and \(g_s\) values we aim to understand the role that orbital and spin angular momenta plays in the connection.
        
        The effect of changing the \(g\) factors can be seen in \cref{fig:gsf_M1_gl_gs_normal}, where three different \(M1\) GSFs of \(^{50}\)V are shown. In red we have \(g_s = (5.027, -3.443)\) and \(g_l = (1.1, -0.1)\). These specific \(g_s\) factors are the free \(g_s\) factors multiplied by a quenching factor of 0.9 as described in \cref{sec:g_values}. For the GSF in dark red, we have turned off the orbital angular momentum contribution to the \(M1\) GSF by setting \(g_l = (0, 0)\), and in orange the spin contribution is removed by setting \(g_s = (0, 0)\).
        
        We see that for all these different combinations of \(g\) factors, the overall \textit{shape} of the \(M1\) GSF remains approximately the same in the LEE energy range of \(E_\gamma = [0, 2]\) MeV. There is however a significant change in the magnitude of the LEE for the different parameter choices. Turning off the spin contribution (orange) and turning off the orbital angular momentum contribution (dark red) has approximately the same effect in the LEE energy range. As both of them are approximately a factor of 5 below the normal \({M}1\) GSF, they seem to be playing a similar role in building up the LEE.
        
        As the gamma energy increases above 2 MeV, we can see in dark red that removing the orbital angular momentum contribution has less and less of an impact, and at the highest energies the impact of its removal is close to negligible. From approximately 6 MeV and up, the \({M}1\) GSF seems to be well described by only the \(\hat{S}\) part of the \(\hat{M}1\) operator.
    
        \begin{figure}[t]
            \centering
            \includegraphics[width=\linewidth]{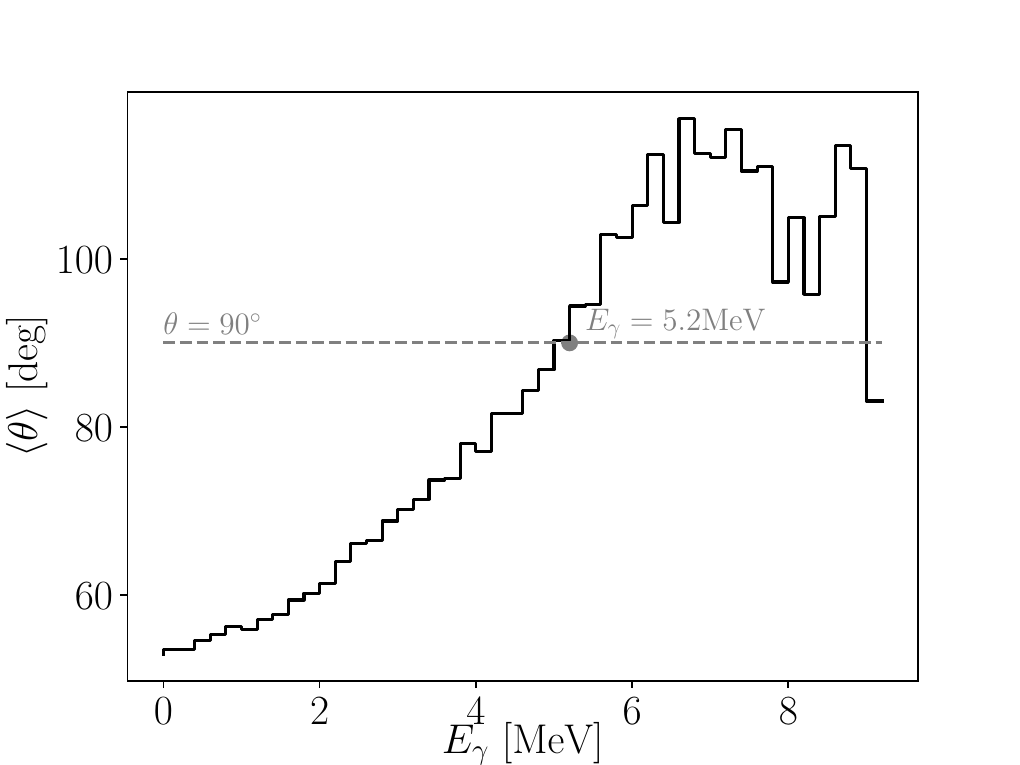}
            \caption{The interference angle \cref{eq:interference-angle} between \(M_l\) and \(M_s\) averaged over the same transitions in the same \(E_\gamma\) bins as in \cref{fig:gsf_M1_gl_gs_normal}. \(\theta = 90^\circ\) is the angle of neither constructive nor destructive interference and happens at \(E_\gamma = 5.2\) MeV.}
            \label{fig:interference-angle}
        \end{figure}
        
        In \cref{fig:interference-angle} we can see the interference angle, as defined in \cref{eq:interference-angle}, averaged over the transitions in each bin of \cref{fig:gsf_M1_gl_gs_normal}. An angle of \(\theta = [0^\circ, 90^\circ)\) means that there is constructive interference, \(90^\circ\) means that there is no interference at all, while \(\theta = (90^\circ, 180^\circ]\) means destructive interference.
        
        We can see that the LEE region of \(E_\gamma = [0, 2]\) MeV has a relatively large constructive interference. This means that summing ``Only \(g_s\)'' and ``Only \(g_l\)'', as shown in green in \cref{fig:gsf_M1_gl_gs_normal}, which excludes the \(2M_lM_s\) term, is \textit{less} than the full \(M1\) strength function; constructive interference makes up for the difference.
        
        As the gamma energy increases, the interference angle follows. At 5.2 MeV the interference angle is at \(90^\circ\) which means that there is no interference at all. This can be seen in \cref{fig:gsf_M1_gl_gs_normal} where the green graph crosses the red; this is the point where the sum of ``Only \(g_s\)'' and ``Only \(g_l\)'' equals the full \(M1\) GSF.
        
        From 5 MeV and beyond the interference angle continues to increase above \(90^\circ\), meaning that \(M_l\) and \(M_s\) are now destructively interfering. This can also be seen from \cref{fig:gsf_M1_gl_gs_normal} where the sum of ``Only \(g_s\)'' and ``Only \(g_l\)'' is larger than the full \(M1\) strength function. At the very highest gamma energies the interference angle starts to decrease, however, because of few transitions in the highest \(E_\gamma\) bins we must be careful about drawing conclusions.

        The interference angle in \cref{fig:interference-angle} clearly shows that constructive interference between orbital and spin angular momentum gives an extra enhancement to the LEE; the enhancement is of approximately a factor of 3. The interference angle steadily increases as \(E_\gamma\) increases, reaches a point of no interference at \(E_\gamma = 5.2\) MeV, and produces a slight destructive interference beyond 5.2 MeV. The destructive interference is however small, and the full \(M1\) GSF is quite well described by just the spin term at these energies.
    
    \subsection{Information from the rOBTDs}

        \begin{figure*}[h!t]
            \centering
            \includegraphics[scale=0.5]{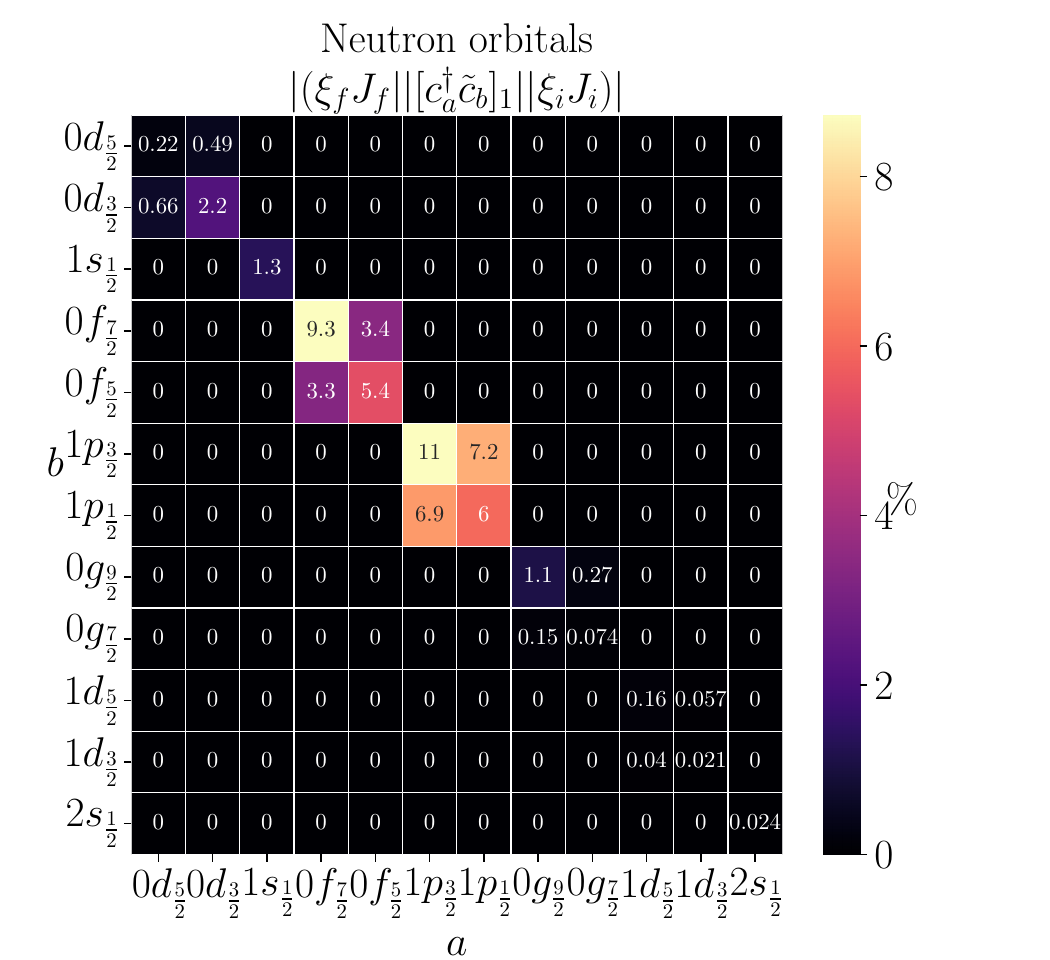}
            \includegraphics[scale=0.5]{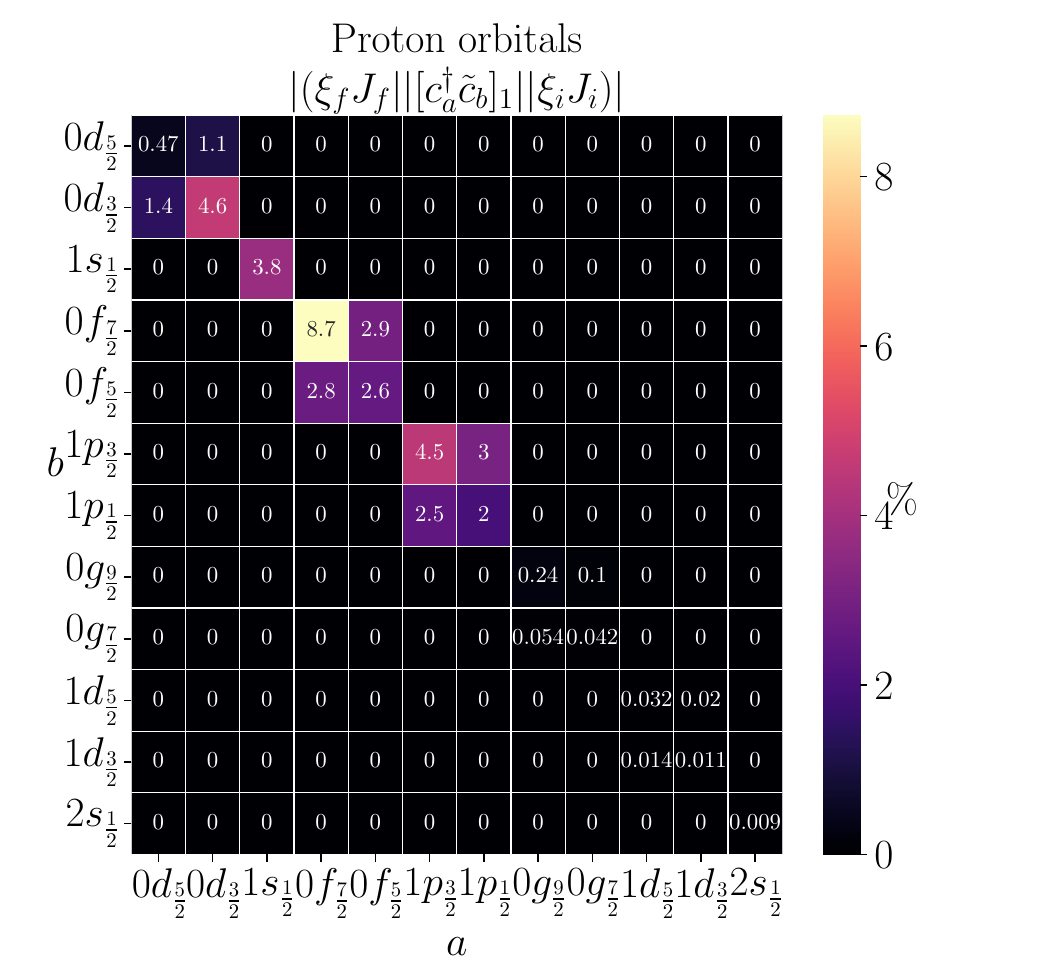}
            \caption{Neutron and proton rOBTDs for all the transitions in the \(E_\gamma = [0, 3]\) MeV interval of the \(M1\) gamma strength function of \(^{50}\)V. The rOBTDs are presented as a percentage of the sum of the absolute value of all the OBTDs in the energy interval (see text).}
            \label{fig:heatmaps}
        \end{figure*}   
        
        We have established that, in the shell-model calculations, the LEE is caused by \(M1\) transitions. Moreover, we have seen that both the orbital and spin angular momentum contribute to the LEE strength. We can pinpoint the origin of the LEE further by considering the reduced one-body transition densities (rOBTD). As defined in \cref{eq:robtd}, the rOBTDs enable us to filter out contributions from specific orientation independent (hence \textit{reduced}) one-particle transitions.
        
        A shell-model wave function is not described as one specific configuration of protons and neutrons in a model space, but rather as a linear combination of many configurations of protons and neutrons (a.k.a. basis states). Consequently, we cannot say that a transition is an exact, specific change in the configuration of the nucleons. With the rOBTDs however, we can see the amount of which the different single-particle transitions contribute to the transition between two wave functions.
    
        We have taken all of the transitions that are contained in the \(E_\gamma = [0, 3]\) MeV LEE region of the \(M1\) GSF, calculated the rOBTDs for each of the transitions, and then summed the \textit{absolute value} of the rOBTDs at each \(a, b\) pair for each of the transitions. Additionally, we have normalised all the rOBTDs to their sum and expressed them as a percentage of the total sum. Since there are 12 proton and 12 neutron single-particle orbitals in the model space, \(a\) and \(b\) will run over all the 12 proton orbitals and all the 12 neutron orbitals, for protons and neutrons separately because a single-particle transition would change the number of protons and neutrons if we let \(a\) and \(b\) run over both proton and neutron orbitals at the same time. Each transition will then have \(12 \times 12\) proton OBTDs and \(12 \times 12\) neutron OBTDs; however, most of the OBTDs will be zero for the following reasons: \(M1\) transitions are only  allowed when there is no change in parity from the initial to the final state. While transitions between \(sd\) and \(sdg\) are technically possible as they do not change the parity, they are made impossible by the \(1 \hbar \omega\) truncation and by the fact that such \(M1\) transitions – where \(n_i \neq n_f\) – are not implemented in the \(\hat{M}1\) operator of the KSHELL code. Therefore, the non-zero \(M1\) rOBTDs will only be \textit{within each major shell}. Additionally, \(M1\) transitions with \(l_i \neq l_f\) are not allowed due to the property of the implemented \(\hat{M}1\) operator \cite{rin80}, meaning that the reduced single-particle matrix element (rSPME)  \(( a \lVert \hat{M}1 \lVert b )\) will be zero. When presenting the rOBTD results in the following, we have  multiplied the rOBTDs with 0 if the accompanying rSPME is zero. This is to avoid cases where we might have a large rOBTD but it would promptly be multiplied by a zero rSPME.
    
        The rOBTD heatmaps are shown in \cref{fig:heatmaps}. Each value in the heatmap is interpreted as how much the single-particle transition from orbital \textit{row} (\(b\)) to orbital \textit{col} (\(a\)) contributes to all of the transitions in the given energy interval.
        
        The neutron orbitals have a slightly larger contribution at \(59\%\) compared to the proton orbitals at \(41 \%\). For both nucleon species, we see that the \(sdg\) major shell has a very small contribution, the largest being \(1.1 \%\) in the neutron \(g_{9/2} \rightarrow g_{9/2}\). The \(sd\) major shell contributes more than the \(sdg\) major shell, and its contribution is greater for protons than for neutrons. The \(pf\) major shell is the largest contributor and contributes more for neutrons than for protons.

        \begin{figure*}[t]
            \centering
            \includegraphics[scale=0.5]{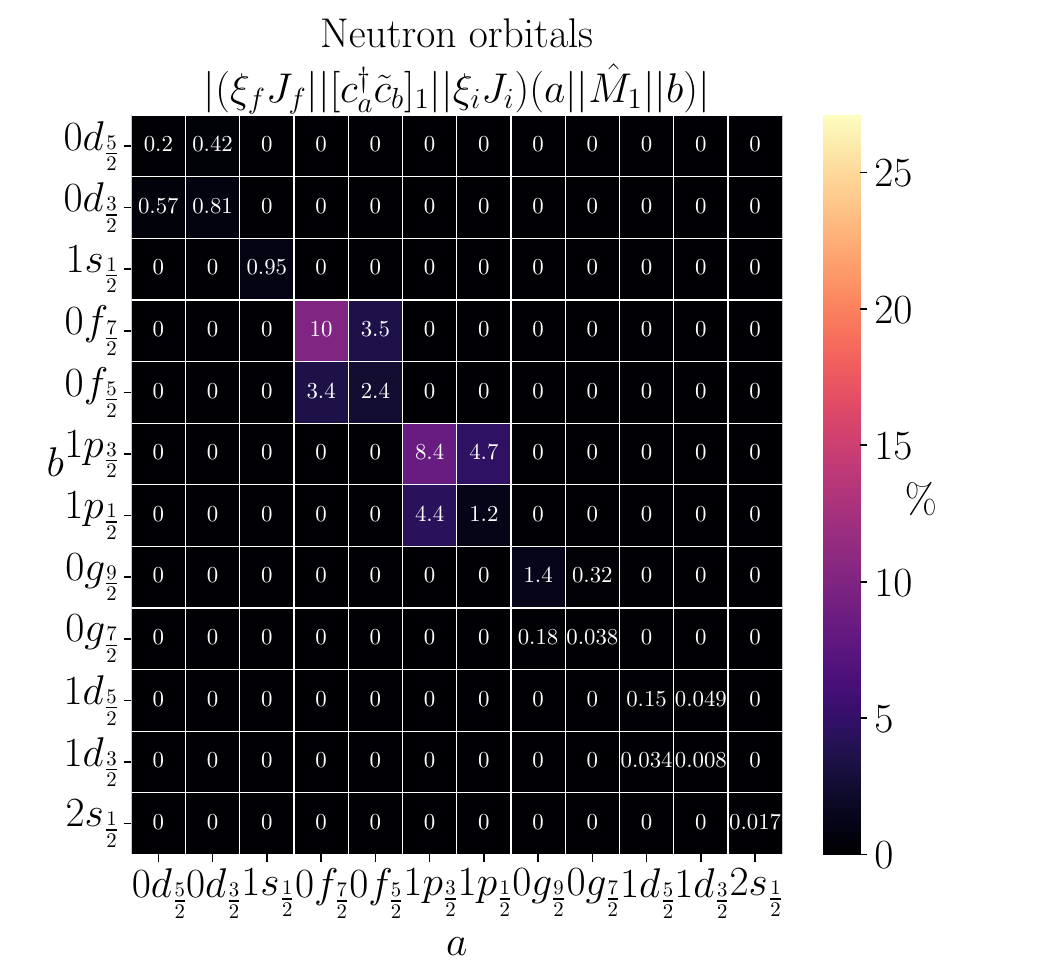}
            \includegraphics[scale=0.5]{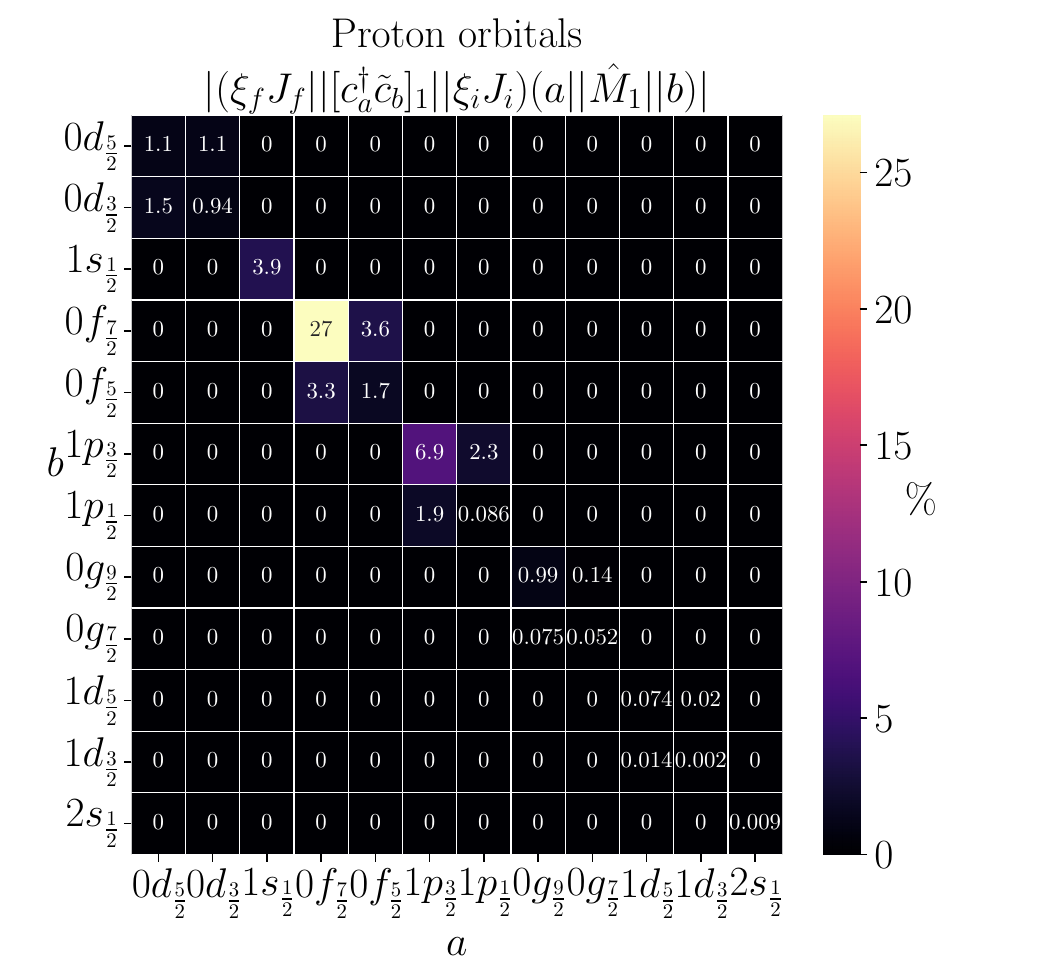}
            \caption{Same as \cref{fig:heatmaps} but the rOBTDs have been multiplied with the accompanying rSPMEs.}
            \label{fig:heatmaps_include_mred}
        \end{figure*}
    
        The general features of the heatmaps in \cref{fig:heatmaps} are as expected. Recall that \(^{50}\)V has 23 protons and 27 neutrons, meaning that in the single-particle picture, the ground state configuration of nucleons will fill up to seven neutrons in \(0f_{7/2}\) and three protons also in \(0_{f7/2}\) (see \cref{fig:nuclear_shell_model}). Since all the valence nucleons initially are in the \(0f_{7/2}\) orbital of the \(pf\) major shell, it is expected that the \(pf\) orbitals will be heavily involved in the dynamics of the valence nucleons simply because the Fermi energy level is in the close proximity of these orbitals. This expectation is confirmed by the rOBTDs in \cref{fig:heatmaps} which shows that the largest contribution comes from the \(pf\) orbitals.

        In the ground-state configuration, both the proton and neutron \(sd\) orbitals are completely filled, prohibiting any permutation of nucleons in the \(sd\) shell. If a nucleon is moved up to the \(pf\) shell – only a single nucleon at a time is allowed to do that because of the \(1 \hbar \omega\) truncation – the parity changes and a hole in the \(sd\) shell is created, allowing nucleon permutations within this shell. If a nucleon is moved from \(pf\) to \(sdg\) the parity also changes, but no hole in \(sd\) is created. Since the natural parity of \isotope[50]{V} is positive, moving one nucleon across a major shell gap will change the parity to negative. The consequence is that the \(M\)-basis states with positive parity \textit{all} have a completely filled \(sd\) major shell. Wave functions of positive parity are exclusively built up of \(M\)-basis states with positive parity, meaning that the \(sd\) rOBTDs will all be zero for positive parity wave-functions, and thus the percentage of rOBTDs in the \(sd\) shell will be lowered relative to the \(pf\) rOBTD percentage. The same argument goes for the \(sdg\) major shell, where in the ground-state configuration there are no nucleons to permute. On the other hand, if we were to allow two nucleons to go from \(sd\) to \(pf\), i.e. a \(2 \hbar \omega\) truncation, it would be possible to create positive-parity basis states, and hence positive-parity wave functions, with various permutations in the \(sd\) major shell, leading to non-zero \(sd\) rOBTDs for the positive-parity states.

        There is a significant amount of non-occupied orbitals between the \(0f_{7/2}\) orbital and the \(sdg\) major shell, putting a large energy-gap between the valence nucleons in the ground state configuration and the \(sdg\) orbitals. Consequently, it is expected that the \(sdg\) orbitals will have a much lower rOBTD contribution than the more energy-accessible \(sd\) and \(pf\) orbitals. In \cref{fig:heatmaps} we indeed see that the \(sdg\) orbitals have a very small rOBTD contribution, however somewhat larger for the neutron case likely because there are more neutrons than protons, meaning that the neutron Fermi level is closer  to the \(sdg\) orbitals are lower for neutrons than the Fermi level for protons.

        While the rOBTDs characterise the many-nucleon properties of the initial and final states, the rSPMEs characterise the one-body operator, in this case the \(\hat{M}1\) operator. If we only want structure information, i.e., which particle–hole channels mix strongly between the two states regardless of operator, then looking at the rOBTD alone is fine. The moment we want to know how much a channel contributes to the electromagnetic transition we need to include the rSPME. In \cref{fig:heatmaps_include_mred} we see the rOBTDs multiplied by the rSPMEs (see \cref{eq:reduced_matrix_element}) and we can immediately see that the orbital contributions have changed. The protons now contribute more, with \(57 \%\), compared to neutrons with \(43 \%\). The proton \(sd\) major shell contribution has slightly decreased from \(11 \% \to 8.5 \%\), while the proton \(pf\) and proton \(sdg\) contributions have increased from \(29 \% \to 47 \%\) and \(0.53 \% \to 1.4 \%\), respectively. The neutron \(pf\) contribution has decreased from \(52 \% \to 38 \%\) with \(sd\) and \(sdg\) approximately unchanged. The majority of the per-orbital contribution now resides in the proton \(0f_{7/2} \to 0f_{7/2}\) with \(27 \%\) of the total.
    
    \subsection{Single-particle orbital contributions to the GSF and LEE}
        While the individual rOBTDs and rSPMEs paint a picture of what is happening at the single-particle level of the shell-model calculations, the information about phase coherence (whether values add up or cancel each other) is lost. We would like to investigate the shape and the amplitude of the LEE as a function of single-particle orbital contributions. Since we have access to all the rOBTDs and the rSPMEs, it is simple to re-calculate any and all transition probabilities by the use of \cref{eq:B} and \cref{eq:matrix_element}, omitting  single-particle orbitals in a systematic way to study their impact.

        \subsubsection{\texorpdfstring{Removing \(0f_{7/2} \rightarrow 0f_{7/2}\)}{Removing 0f7/2 to 0f7/2}}
            \begin{figure*}[t]
                \centering
                \includegraphics[scale=0.5]{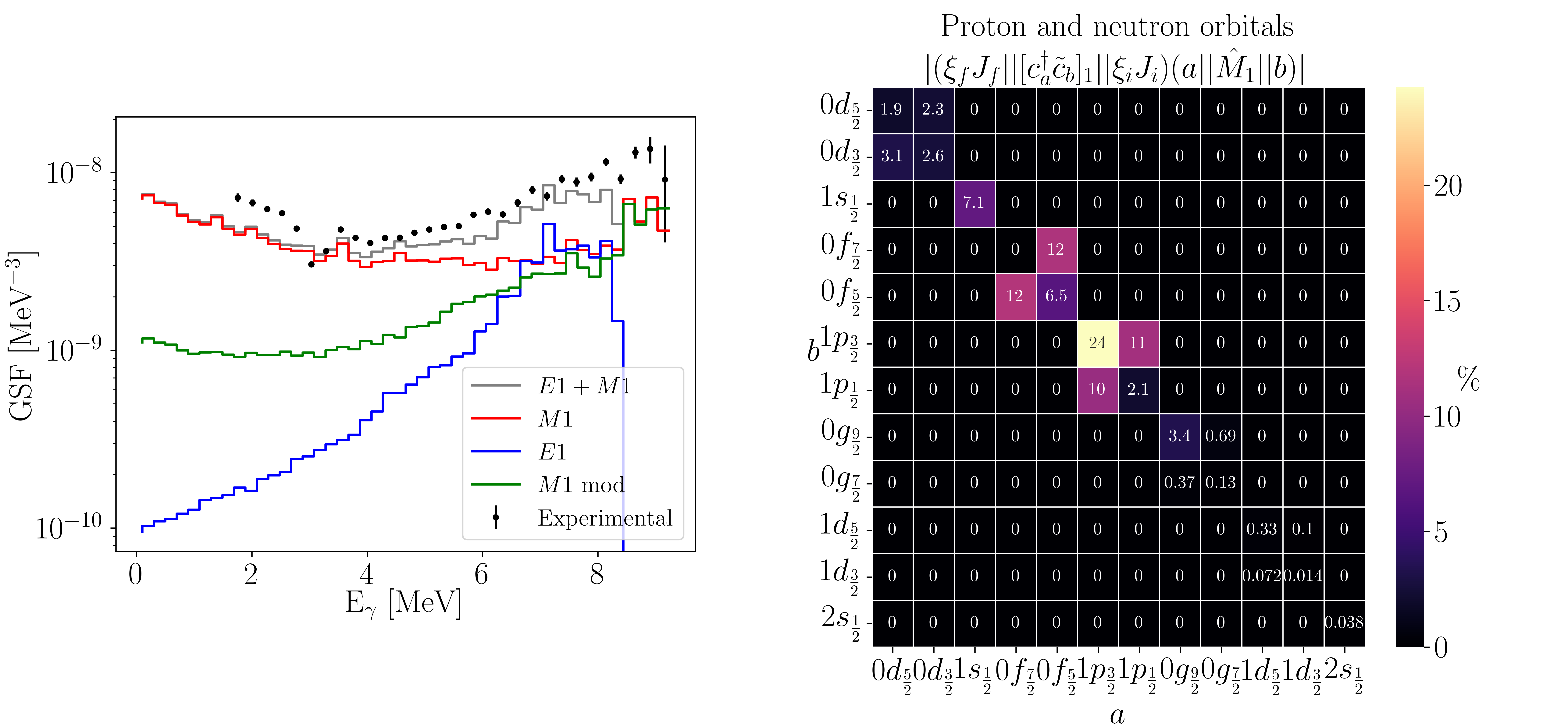}
                \caption{Left: The re-calculated \(M1\) strength function in green. Right: Accompanying map of the rOBTDs \(\times\) rSPMEs summed for protons and neutrons, showing that \(0f_{7/2} \rightarrow 0f_{7/2}\) has been completely removed.}
                \label{fig:gsf_recalc_remove_f7f7}
            \end{figure*}
        
            In \cref{fig:gsf_recalc_remove_f7f7}, we display the re-calculated \(M1\) strength function in green, where the contribution from \(0f_{7/2} \rightarrow 0f_{7/2}\) has been completely removed. 
            The impact is substantial, and it is largest in the low-energy region, having reduced the original \(M1\) strength by approximately a factor of 7. The impact lessens as the gamma energy approaches \(\approx 7\) MeV where the modified and original \(M1\) strength functions meet. The LEE is almost eliminated with only a tiny increase left over at the very lowest gamma energies. By removing \(0f_{7/2} \rightarrow 0f_{7/2}\) we see that most of the rOBTD weight shifts over to \(1p_{3/2} \rightarrow 1p_{3/2}\).
    
        \subsubsection{\texorpdfstring{Removing  \(1p_{3/2} \rightarrow 1p_{3/2}\)}{Removing  1p3/2 to 1p3/2}}
            \begin{figure*}[t]
                \centering
                \includegraphics[scale=0.5]{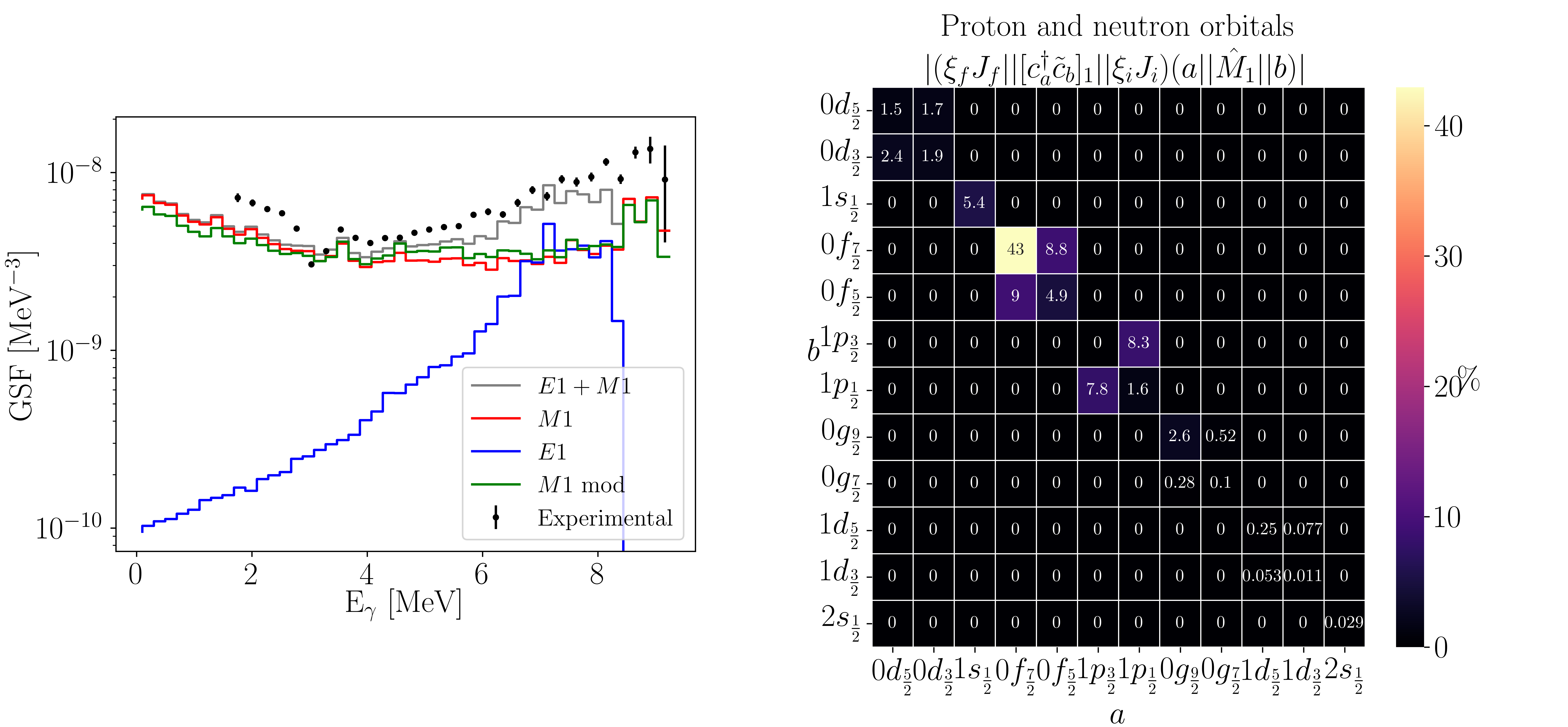}
                \caption{Left: The re-calculated \(M1\) strength function in green. Right: Accompanying map of the rOBTDs \(\times\) rSPMEs summed for protons and neutrons, showing that \(1p_{3/2} \rightarrow 1p_{3/2}\) has been completely removed.}
                \label{fig:gsf_recalc_remove_p3p3}
            \end{figure*}
        
            Completely removing the contribution from \(1p_{3/2} \rightarrow 1p_{3/2}\), as seen in \cref{fig:gsf_recalc_remove_p3p3}, does not have much of an impact even though it contributes to a large fraction of the total OBTDs. The \(M1\) strength function is only slightly changed by removing its contribution.
    
        \subsubsection{Removing the diagonal contributions}
            \begin{figure*}[t]
                \centering
                \includegraphics[scale=0.5]{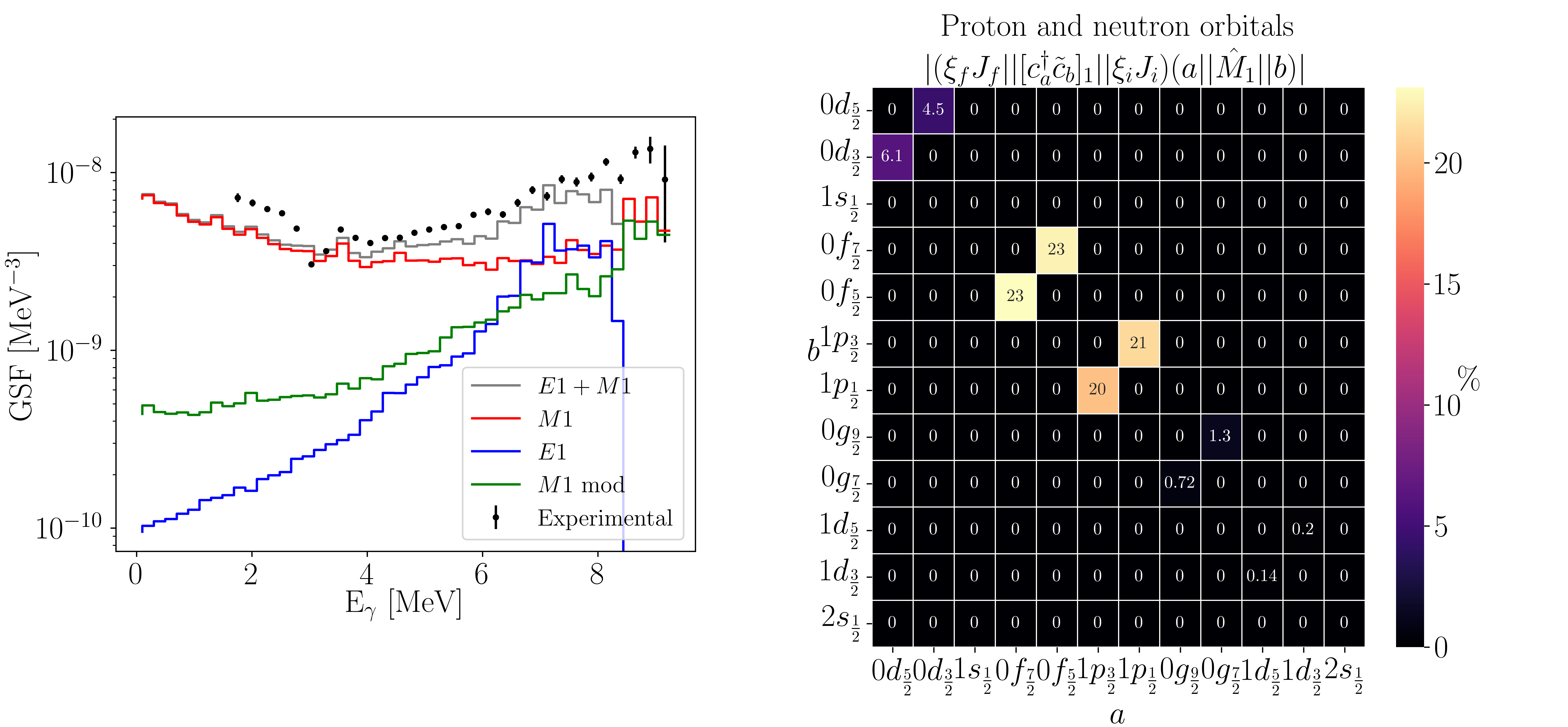}
                \caption{Left: The re-calculated \(M1\) strength function in green. Right: Accompanying map of the rOBTDs \(\times\) rSPMEs summed for protons and neutrons, showing that all the diagonal contributions have been completely removed.}
                \label{fig:gsf_recalc_remove_diagonal}
            \end{figure*}
        
            In \cref{fig:gsf_recalc_remove_diagonal} all of the diagonal contributions have been completely removed and most of the low-energy strength has gone with them. Any trace of a LEE is completely gone. The difference to \cref{fig:gsf_recalc_remove_f7f7} is however not very large, indicating again that most of the low-energy strength comes from \(0f_{7/2}\rightarrow 0f_{7/2}\) transitions.
    
        \subsubsection{Removing the non-diagonal contributions}
            \begin{figure*}[t]
                \centering
                \includegraphics[scale=0.5]{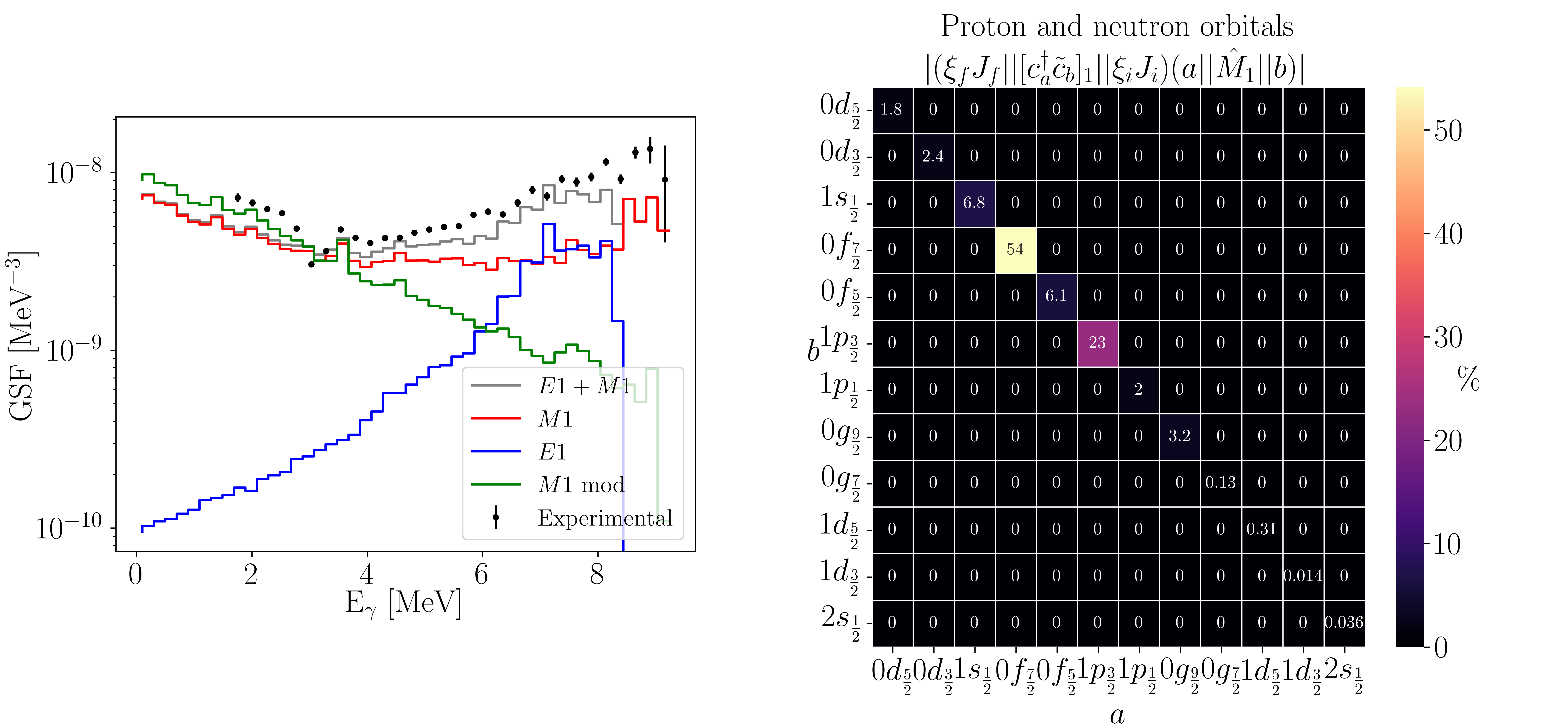}
                \caption{Left: The re-calculated \(M1\) strength function in green. Right: Accompanying map of the rOBTDs \(\times\) rSPMEs summed for protons and neutrons, showing that all the non-diagonal contributions have been completely removed.}
                \label{fig:gsf_recalc_remove_nondiagonal}
            \end{figure*}
        
            All the non-diagonal contributions have been removed in \cref{fig:gsf_recalc_remove_nondiagonal}, and the effects are the opposite of what we saw when removing the diagonal contributions. The low-energy strength is kept, and is actually a bit higher than the original \(M1\) strength function, indicating that the off-diagonal contributions are actually pulling the low-energy strength down slightly. At the approximate mid-point of $E_\gamma \approx 4$ MeV, however, the strength falls compared to the original, and keeps falling steadily until the highest gamma energies are reached. It seems conclusive that the low-energy strength is contained in the diagonal elements, and by far mainly in the \(0f_{7/2} \rightarrow 0f_{7/2}\) transitions in accordance with the findings of Refs.~\cite{Brown2014,PhysRevC.95.024322}, while the high-energy strength, i.e. \(E_\gamma > 4\) MeV, is contained in the off-diagonal elements. This seems to be in accordance with the fact that the spin-flip \(M1\) transitions dominate for higher energy gamma-ray transitions, as shown experimentally for the neighbouring nucleus $^{51}$V~\cite{DJALALI19821}.
\section{Summary and outlook}\label{seq:summary}
    We have carried out the first shell–model investigation of \isotope[50]{V} in a valence space that spans the three major shells \(sd\), \(pf\) and \(sdg\).  We have used the KSHELL code, whose thick-restart block-Lanczos algorithm efficiently solves the eigenvalue problem.  With the \texttt{SDPFSDG-MU} interaction and a \(1\hbar\omega\) truncation we handled \(M\)-scheme bases of \(7.02\times10^{6}\) states for positive parity and \(5.94\times10^{8}\) states for negative parity, yielding nearly two million individual \(E1\) and \(M1\) transition strengths—an exceptionally large data set for shell-model calculations.
    
    Benchmark comparisons with experimental data demonstrate the quality of the calculations.  The fourteen lowest observed states are reproduced within \(0.30\;\text{MeV}\) and the total level density follows Oslo-method data up to \(E_x\!\approx\!7.5\;\text{MeV}\).  The dipole gamma strength function calculated from these levels reproduces the experimental shape from \(E_\gamma\!\approx\!1\) to \(7\;\text{MeV}\), including the low-energy enhancement. The difference in absolute magnitude is comparable to the experimental normalisation uncertainty, which stems from the unmeasured total average radiative width for \isotope[50]{V}.
    
    With both the $E1$ and $M1$ strength functions calculated within the same theoretical framework, we conclude that the LEE is entirely magnetic \emph{dipole} in origin for \isotope[50]{V}. Tests in which the spin or orbital parts of the \(\hat{M}1\) operator are suppressed show that both components are necessary to reproduce the enhancement. The spin term is the most important one beyond \(E_\gamma\!\sim\!2\;\text{MeV}\) and the orbital contribution becomes negligible above \(5\;\text{MeV}\). By investigating the interference angle between the orbital and spin contribution to the \(\hat{M}1\) operator, we see that constructive interference gives an extra enhancement to the LEE of approximately a factor of 3. The interference angle has proved to give valuable information about the \(M1\) strength.
    
    Our analysis of reduced one-body transition densities identifies diagonal \(0f_{7/2}\!\rightarrow\!0f_{7/2}\) proton transitions as the principal source; removing this single channel lowers the low-energy \(M1\) strength by approximately a factor of 7, whereas off-diagonal particle–hole excitations govern the spectrum for \(E_\gamma\!\gtrsim\!4\;\text{MeV}\).
    
    Because our calculations provide an exceptionally large collection of levels and transitions, it now becomes possible to perform rigorous statistical studies—most notably tests of Porter–Thomas fluctuations and of the generalised Brink–Axel hypothesis.  That analysis is under way and will be presented in a forthcoming work.  A natural theoretical extension is to apply the same \(sd\text{--}pf\text{--}sdg\) framework to neighbouring vanadium isotopes as well as other nuclei in this mass region, thereby mapping the systematics of the \(0f_{7/2}\)-driven enhancement.
    
    By linking detailed shell-model structure to dipole radiation over a vast configuration space, the present work demonstrates a quantitative route from microscopic configurations to emergent statistical properties—an essential step toward a unified description of nuclear structure and reactions.

\begin{acknowledgments}
    The calculations were performed on resources provided by Sigma2, the National Infrastructure for High Performance Computing and Data Storage in Norway using ``Betzy'' on Project No. NN9464K. J.~K.~D. and A.~C.~L. gratefully acknowledge continued support from the Centre for Computational and Data Science (dScience) at the University of Oslo, Norway. J.~K.~D. and A.~C.~L. acknowledge financial support from the Research Council of Norway, Project No. 316116, and support from the Norwegian Nuclear Research Centre, Project No. 341985. N. S. and Y. U. acknowledge the support of the ``Program for promoting research on the supercomputer Fugaku'', MEXT, Japan (JPMXP1020230411), JST ERATO Grant No. JPMJER2304, Japan, and KAKENHI (25K00995 and 25K07330). The authors sincerely thank M.~Hjorth-Jensen and J.~A.~Fløisand for stimulating discussions.
\end{acknowledgments}

The data that support the findings in this article are openly available at \cite{DahlZenodo2026}.

\onecolumngrid
\newpage
\appendix
    \section{Deriving the reduced transition probability} \label{sec:red-trans-prob}
    In this appendix we derive the reduced matrix element
    \begin{align} \label{eq:app-reduced-matrix-element}
        (J_f& \lVert \hat{\bm{O}}_\lambda \lVert J_i ) = \widehat{\lambda}^{-1} \sum_{a b} ( a \lVert \hat{\bm{O}}_\lambda \lVert b ) (  J_f \lVert [c_a^{\dagger} \tilde{c}_b]_\lambda \lVert J_i),
    \end{align}
    from which we calculate the reduced transition probability \(B\).
    
    In occupation number representation, a one-body operator \(\hat{O}_{\lambda \mu}\) can be represented by
    \begin{align} \label{eq:app-occupation-number-representation}
        \hat{O}_{\lambda \mu} = \sum_{\alpha \beta} \langle \alpha | \hat{O}_{\lambda \mu} | \beta \rangle \hat{c}_\alpha^\dagger \hat{c}_\beta
    \end{align}
    where \(\alpha\) (\(\beta\)) is short-hand for the quantum numbers needed to describe a specific \(m\)-substate, \(| \alpha \rangle \equiv | j_\alpha m_\alpha \rangle\), and \(\hat{c}\), \(\hat{c}^\dagger\) are the annihilation and creation operators respectively. Thus, the transition matrix element is given by
    \begin{align} \label{eq:app-occupation-number-representation-with-sandwich}
        \langle J_f M_f | \hat{O}_{\lambda \mu} | J_i M_i \rangle = \langle J_f M_f | \left( \sum_{\alpha \beta} \langle \alpha | \hat{O}_{\lambda \mu} | \beta \rangle \hat{c}_\alpha^\dagger \hat{c}_\beta \right) | J_i M_i \rangle.
    \end{align}
    The goal is to use the Wigner-Eckhart theorem on both sides of \cref{eq:app-occupation-number-representation-with-sandwich} to show the equality in \cref{eq:app-reduced-matrix-element}.

    Assume that \(\hat{\bm{O}}_{\lambda}\) is a spherical tensor operator of rank \(\lambda\). Since \(\hat{\bm{O}}_\lambda\) is a spherical tensor operator, the Wigner-Eckhart theorem tells us that the l.h.s. of \cref{eq:app-occupation-number-representation-with-sandwich} is given by
    \begin{align} \label{eq:app-wigner-eckart-rewrite}
        \langle J_f M_f | \hat{O}_{\lambda \mu} | J_i M_i \rangle = \dfrac{1}{\sqrt{2 J_f + 1}} ( J_f \lVert \hat{\bm{O}}_\lambda \lVert J_i ) ( J_i \; M_i \; \lambda \; \mu \mid J_f \; M_f ).
    \end{align}
    On the r.h.s. of \cref{eq:app-occupation-number-representation-with-sandwich} we have two factors which separately needs to be put through the Wigner-Eckhart machinery, namely the matrix element and the creation-annihilation pair. The Wigner-Eckhart theorem applied to the matrix element gives
    \begin{align} \label{eq:app-wigner-eckart-rewrite-2}
        \langle j_\alpha m_\alpha | \hat{O}_{\lambda \mu} | j_\beta m_\beta \rangle = \dfrac{1}{\sqrt{2 j_\alpha + 1}} ( j_\alpha \lVert \hat{\bm{O}}_\lambda \lVert j_\beta ) ( j_\beta \; m_\beta \; \lambda \; \mu \mid j_\alpha \; m_\alpha ).
    \end{align}
    The creation-annihilation pair is however not a spherical tensor operator so we have to convert them into one before we can use the Wigner-Eckhart theorem. We know that if \(\bm{T}_{L_1}\) and \(\bm{T}_{L_2}\) are two spherical tensor operators of rank \(L_1\) and \(L_2\) respectively, then their tensor product – which is also a spherical tensor operator – is denoted \(\bm{T}_{L}\) whose \(2L + 1\) components are defined as (see for example \cite[sec. 2.3]{suhonen})
    \begin{align}
        T_{L M} = \sum_{M_1 M_2}\left(L_1 \; M_1 \; L_2 \; M_2 \mid L \; M\right) T_{L_1 M_1} T_{L_2 M_2} \equiv\left[\bm{T}_{L_1} \bm{T}_{L_2}\right]_{L M}.
    \end{align}
    The annihilation operator \(\hat{c}_\beta\) is not a spherical tensor operator, however
    \begin{align}
        \tilde{c}_\beta \equiv(-1)^{j_\beta + m_\beta} \hat{c}_{j_\beta, -m_\beta}
    \end{align}
    is. Looking back at the r.h.s. of \cref{eq:app-occupation-number-representation-with-sandwich}, we insert \cref{eq:app-wigner-eckart-rewrite-2} to get
    \begin{align} \label{eq:app-intermediate-term-1}
        \langle J_f M_f | \left( \sum_{\alpha \beta} \langle \alpha | \hat{O}_{\lambda \mu} | \beta \rangle \hat{c}_\alpha^\dagger \hat{c}_\beta \right) | J_i M_i \rangle = \langle J_f M_f | \left( \sum_{\alpha \beta} \dfrac{1}{\sqrt{2 j_\alpha + 1}} ( j_\alpha \lVert \hat{\bm{O}}_\lambda \lVert j_\beta ) ( j_\beta \; m_\beta \; \lambda \; \mu \mid j_\alpha \; m_\alpha ) \hat{c}_\alpha^\dagger \hat{c}_\beta \right) | J_i M_i \rangle.
    \end{align}
    The keen-eyed of you might see the similarities between
    \begin{align}
        \sum_{\alpha \beta} ( j_\beta \; m_\beta \; \lambda \; \mu \mid j_\alpha \; m_\alpha ) \hat{c}_\alpha^\dagger \hat{c}_\beta
        && \text{and} &&
        \sum_{M_1 M_2}\left(L_1 \; M_1 \; L_2 \; M_2 \mid L \; M\right) T_{L_1 M_1} T_{L_2 M_2},
    \end{align}
    however with a different order on the labels in the Clebsch-Gordan (CG) coefficients. One of the symmetry properties of the CG coefficients is
    \begin{align} \label{eq:app-cg-symmetry-property}
        ( j_\alpha \; m_\alpha \; j_\beta \; m_\beta \mid \lambda \; \mu ) = (-1)^{j_\beta + m_\beta} \sqrt{\frac{2 \lambda + 1}{2 j_\alpha + 1}}\left( j_\beta \; \left(-m_\beta\right) \; \lambda \; \mu \mid j_\alpha \; m_\alpha\right),
    \end{align}
    which we can use to bridge the gap. There is however a \(-m_\beta\) in \cref{eq:app-cg-symmetry-property} which does not appear in \cref{eq:app-intermediate-term-1}. This can be solved by reversing the order of summation in the \(m_\beta\) sum, which is perfectly fine to do as addition is commutative. We start by separating the \(j\) and \(m\) sums:
    \begin{align}
        \sum_{j_\alpha j_\beta} \sum_{m_\alpha m_\beta}& \dfrac{1}{\sqrt{2 j_\alpha + 1}} ( j_\alpha \lVert \hat{\bm{O}}_\lambda \lVert j_\beta ) ( j_\beta \; m_\beta \; \lambda \; \mu \mid j_\alpha \; m_\alpha ) \hat{c}_{j_\alpha m_\alpha}^\dagger \hat{c}_{j_\beta m_\beta}
        \\
        &= \sum_{j_\alpha j_\beta} \sum_{m_\alpha m_\beta} \dfrac{1}{\sqrt{2 j_\alpha + 1}} ( j_\alpha \lVert \hat{\bm{O}}_\lambda \lVert j_\beta ) ( j_\beta \; (-m_\beta) \; \lambda \; \mu \mid j_\alpha \; m_\alpha ) \hat{c}_{j_\alpha m_\alpha}^\dagger \hat{c}_{j_\beta (-m_\beta)}
        \\
        &= \sum_{j_\alpha j_\beta} \sum_{m_\alpha m_\beta} \dfrac{1}{\sqrt{2 j_\alpha + 1}} ( j_\alpha \lVert \hat{\bm{O}}_\lambda \lVert j_\beta ) (-1)^{j_\beta + m_\beta} \dfrac{\sqrt{2j_\alpha + 1}}{\sqrt{2 \lambda + 1}} ( j_\alpha \; m_\alpha \; j_\beta \; m_\beta \mid \lambda \; \mu ) \hat{c}_{j_\alpha m_\alpha}^\dagger \hat{c}_{j_\beta (-m_\beta)}
        \\
        &= \dfrac{1}{\sqrt{2 \lambda + 1}} \sum_{j_\alpha j_\beta} ( j_\alpha \lVert \hat{\bm{O}}_\lambda \lVert j_\beta ) \sum_{m_\alpha m_\beta} ( j_\alpha \; m_\alpha \; j_\beta \; m_\beta \mid \lambda \; \mu ) \hat{c}_{j_\alpha m_\alpha}^\dagger (-1)^{j_\beta + m_\beta} \hat{c}_{j_\beta (-m_\beta)}
        \\  \label{eq:app-intermediate-term-2}
        &= \hat{\lambda}^{-1} \sum_{a b} ( a \lVert \hat{\bm{O}}_\lambda \lVert b ) [ \hat{c}_{a}^\dagger \tilde{c}_{b}]_{\lambda \mu}
    \end{align}
    Note that \(a\) (\(b\)) is short-hand for the quantum numbers needed to describe a specific \(j\)-state without specifying the orientation: \(| a \rangle \equiv | j_\alpha \rangle\). We tack on initial and final states and use the Wigner-Eckhart theorem on \cref{eq:app-intermediate-term-2} to get
    \begin{align} \label{eq:app-wigner-eckhart-rewrite-rhs}
        \hat{\lambda}^{-1} \sum_{a b} ( a \lVert \hat{\bm{O}}_\lambda \lVert b ) \langle J_f M_f | [ \hat{c}_{a}^\dagger \tilde{c}_{b}]_{\lambda \mu} |J_i  M_i \rangle = \hat{\lambda}^{-1} \sum_{a b} ( a \lVert \hat{\bm{O}}_\lambda \lVert b ) \dfrac{1}{\sqrt{2 J_f + 1}} (J_i \; M_i \; \lambda \; \mu  \mid J_f \; M_f ) ( J_f \lVert [ \hat{c}_{a}^\dagger \tilde{c}_{b}]_{\lambda} \lVert J_i).
    \end{align}
    Finally putting together the l.h.s. and r.h.s., ie. \cref{eq:app-wigner-eckart-rewrite} and \cref{eq:app-wigner-eckhart-rewrite-rhs}, we get
    \begin{align}
        \dfrac{1}{\sqrt{2 J_f + 1}} ( J_f \lVert \hat{\bm{O}}_\lambda \lVert J_i ) ( J_i \; M_i \; \lambda \; \mu | J_f \; M_f ) &= \hat{\lambda}^{-1} \sum_{a b} ( a \lVert \hat{\bm{O}}_\lambda \lVert b ) \dfrac{1}{\sqrt{2 J_f + 1}} (J_i \; M_i \; \lambda \; \mu  \mid J_f \; M_f ) ( J_f \lVert [ \hat{c}_{a}^\dagger \tilde{c}_{b}]_{\lambda} \lVert J_i),
        \\
        ( J_f \lVert \hat{\bm{O}}_\lambda \lVert J_i ) &= \hat{\lambda}^{-1} \sum_{a b} ( a \lVert \hat{\bm{O}}_\lambda \lVert b ) ( J_f \lVert [ \hat{c}_{a}^\dagger \tilde{c}_{b}]_{\lambda} \lVert J_i).
    \end{align}

\newpage
\onecolumngrid
\bibliography{references.bib}

@article{KSHELL,
    title    = {Thick-restart block Lanczos method for large-scale shell-model calculations},
    journal  = {Computer Physics Communications},
    volume   = {244},
    pages    = {372-384},
    year     = {2019},
    doi      = {10.1016/j.cpc.2019.06.011},
    author   = {Shimizu, Noritaka and Mizusaki, Takahiro and Utsuno, Yutaka and Tsunoda, Yusuke},
}

@article{sdpfmu,
  title     = {Shape transitions in exotic Si and S isotopes and tensor-force-driven Jahn-Teller effect},
  journal   = {Phys. Rev. C},
  volume    = {86},
  issue     = {5},
  pages     = {051301},
  year      = {2012},
  doi       = {10.1103/PhysRevC.86.051301},
  author    = {Utsuno, Yutaka and Otsuka, Takaharu and Brown, B. Alex and Honma, Michio and Mizusaki, Takahiro and Shimizu, Noritaka},
}

@article{PhysRevC.73.064301,
  title     = {Microcanonical entropies and radiative strength functions of $^{50,51}\mathrm{V}$},
  journal   = {Phys. Rev. C},
  volume    = {73},
  issue     = {6},
  pages     = {064301},
  year      = {2006},
  doi       = {10.1103/PhysRevC.73.064301},
  author    = {Larsen, A. C. and Chankova, R. and Guttormsen, M. and Ingebretsen, F. and Messelt, S. and Rekstad, J. and Siem, S. and Syed, N. U. H. and \O{}deg\aa{}rd, S. W. and L\"onnroth, T. and Schiller, A. and Voinov, A.},
}

@phdthesis{tavukcu_phd,
  author  = {Tavukcu, Emel},
  title   = {Level densities and radiative strength functions in $^{56}$Fe and $^{57}$Fe},
  school  = {North Carolina State University},
  address = {Raleigh, NC},
  year    = {2002}
}

@article{PhysRevC.71.044307,
  title = {Radiative strength functions in $^{93\ensuremath{-}98}\mathrm{Mo}$},
  journal = {Phys. Rev. C},
  volume = {71},
  issue = {4},
  pages = {044307},
  year = {2005},
  month = {Apr},
  doi = {10.1103/PhysRevC.71.044307},
  author = {Guttormsen, M. and Chankova, R. and Agvaanluvsan, U. and Algin, E. and Bernstein, L. A. and Ingebretsen, F. and L\"onnroth, T. and Messelt, S. and Mitchell, G. E. and Rekstad, J. and Schiller, A. and Siem, S. and Sunde, A. C. and Voinov, A. and \O{}deg\aa{}rd, S.},
}

@article{PhysRevLett.118.092502,
  title = {Low-Energy Magnetic Dipole Radiation in Open-Shell Nuclei},
  author = {Schwengner, R. and Frauendorf, S. and Brown, B. A.},
  journal = {Phys. Rev. Lett.},
  volume = {118},
  issue = {9},
  pages = {092502},
  year = {2017},
  month = {Mar},
  doi = {10.1103/PhysRevLett.118.092502},
}

@article{PhysRevC.93.034303,
  title = {First observation of low-energy $\ensuremath{\gamma}$-ray enhancement in the rare-earth region},
  author = {Simon, A. and Guttormsen, M. and Larsen, A. C. and Beausang, C. W. and Humby, P. and Harke, J. T. and Casperson, R. J. and Hughes, R. O. and Ross, T. J. and Allmond, J. M. and Chyzh, R. and Dag, M. and Koglin, J. and McCleskey, E. and McCleskey, M. and Ota, S. and Saastamoinen, A.},
  journal = {Phys. Rev. C},
  volume = {93},
  issue = {3},
  pages = {034303},
  year = {2016},
  month = {Mar},
  doi = {10.1103/PhysRevC.93.034303},
}

@article{PhysRevLett.93.142504,
  title = {Large Enhancement of Radiative Strength for Soft Transitions in the Quasicontinuum},
  author = {Voinov, A. and Algin, E. and Agvaanluvsan, U. and Belgya, T. and Chankova, R. and Guttormsen, M. and Mitchell, G. E. and Rekstad, J. and Schiller, A. and Siem, S.},
  journal = {Phys. Rev. Lett.},
  volume = {93},
  issue = {14},
  pages = {142504},
  year = {2004},
  month = {Sep},
  doi = {10.1103/PhysRevLett.93.142504},
}

@article{PhysRevLett.108.162503,
  title = {Low-Energy Enhancement in the Photon Strength of $^{95}\mathrm{Mo}$},
  author = {Wiedeking, M. and Bernstein, L. A. and Krti\ifmmode \check{c}\else \v{c}\fi{}ka, M. and Bleuel, D. L. and Allmond, J. M. and Basunia, M. S. and Harke, J. T. and Fallon, P. and Firestone, R. B. and Goldblum, B. L. and Hatarik, R. and Lake, P. T. and Lee, I-Y. and Lesher, S. R. and Paschalis, S. and Petri, M. and Phair, L. and Scielzo, N. D.},
  journal = {Phys. Rev. Lett.},
  volume = {108},
  pages = {162503},
  numpages = {5},
  year = {2012},
  month = {Apr},
  doi = {10.1103/PhysRevLett.108.162503},
}

@article{PhysRevLett.111.242504,
  title = {Evidence for the Dipole Nature of the Low-Energy $\ensuremath{\gamma}$ Enhancement in $^{56}\mathrm{Fe}$},
  author = {Larsen, A. C. and Blasi, N. and Bracco, A. and Camera, F. and Eriksen, T. K. and G\"orgen, A. and Guttormsen, M. and Hagen, T. W. and Leoni, S. and Million, B. and Nyhus, H. T. and Renstr\o{}m, T. and Rose, S. J. and Ruud, I. E. and Siem, S. and Tornyi, T. and Tveten, G. M. and Voinov, A. V. and Wiedeking, M.},
  journal = {Phys. Rev. Lett.},
  volume = {111},
  issue = {24},
  pages = {242504},
  year = {2013},
  month = {Dec},
  doi = {10.1103/PhysRevLett.111.242504},
}

@article{Larsen_2017,
  author  = {Larsen, A. C. and Guttormsen, M. and Blasi, N. and Bracco, A. and Camera, F. and Crespo Campo, L. and Eriksen, T. K. and Görgen, A. and Hagen, T. W. and Ingeberg, V. W. and Kheswa, B. V. and Leoni, S. and Midtbø, J. E. and Million, B. and Nyhus, H. T. and Renstrøm, T. and Rose, S. J. and Ruud, I. E. and Siem, S. and Tornyi, T. G. and Tveten, G. M. and Voinov, A. V. and Wiedeking, M. and Zeiser, F.},
  title   = {Low-energy enhancement and fluctuations of $\gamma$-ray strength functions in $^{56,57}$Fe: test of the Brink--Axel hypothesis},
  journal = {J. Phys. G: Nucl. Part. Phys.},
  volume  = {44},
  number  = {6},
  pages   = {064005},
  year    = {2017},
  doi     = {10.1088/1361-6471/aa644a}
}

@article{PhysRevC.97.024327,
  title = {Examination of the low-energy enhancement of the $\ensuremath{\gamma}$-ray strength function of $^{56}\mathrm{Fe}$},
  author = {Jones, M. D. and Macchiavelli, A. O. and Wiedeking, M. and Bernstein, L. A. and Crawford, H. L. and Campbell, C. M. and Clark, R. M. and Cromaz, M. and Fallon, P. and Lee, I. Y. and Salathe, M. and Wiens, A. and Ayangeakaa, A. D. and Bleuel, D. L. and Bottoni, S. and Carpenter, M. P. and Davids, H. M. and Elson, J. and G\"orgen, A. and Guttormsen, M. and Janssens, R. V. F. and Kinnison, J. E. and Kirsch, L. and Larsen, A. C. and Lauritsen, T. and Reviol, W. and Sarantites, D. G. and Siem, S. and Voinov, A. V. and Zhu, S.},
  journal = {Phys. Rev. C},
  volume = {97},
  issue = {2},
  pages = {024327},
  numpages = {6},
  year = {2018},
  month = {Feb},
  publisher = {American Physical Society},
  doi = {10.1103/PhysRevC.97.024327},
  url = {https://link.aps.org/doi/10.1103/PhysRevC.97.024327}
}

@article{PhysRevC.81.024319,
  title = {$\ensuremath{\gamma}$-strength functions in $^{60}\mathrm{Ni}$ from two-step cascades following proton capture},
  author = {Voinov, A. and Grimes, S. M. and Brune, C. R. and Guttormsen, M. and Larsen, A. C. and Massey, T. N. and Schiller, A. and Siem, S.},
  journal = {Phys. Rev. C},
  volume = {81},
  issue = {2},
  pages = {024319},
  numpages = {7},
  year = {2010},
  month = {Feb},
  publisher = {American Physical Society},
  doi = {10.1103/PhysRevC.81.024319},
  url = {https://link.aps.org/doi/10.1103/PhysRevC.81.024319}
}

@article{PhysRevLett.111.232504,
  title = {Low-Energy Enhancement of Magnetic Dipole Radiation},
  author = {Schwengner, R. and Frauendorf, S. and Larsen, A. C.},
  journal = {Phys. Rev. Lett.},
  volume = {111},
  issue = {23},
  pages = {232504},
  numpages = {5},
  year = {2013},
  month = {Dec},
  publisher = {American Physical Society},
  doi = {10.1103/PhysRevLett.111.232504},
  url = {https://link.aps.org/doi/10.1103/PhysRevLett.111.232504}
}

@article{PhysRevC.77.054319,
  title = {Two-step \ensuremath{\gamma} cascades following thermal neutron capture in $^{95}\mathrm{Mo}$},
  author = {Krti\ifmmode \check{c}\else \v{c}\fi{}ka, M. and Be\ifmmode \check{c}\else \v{c}\fi{}v\'a\ifmmode \check{r}\else \v{r}\fi{}, F. and Tomandl, I. and Rusev, G. and Agvaanluvsan, U. and Mitchell, G. E.},
  journal = {Phys. Rev. C},
  volume = {77},
  issue = {5},
  pages = {054319},
  numpages = {15},
  year = {2008},
  month = {May},
  publisher = {American Physical Society},
  doi = {10.1103/PhysRevC.77.054319},
  url = {https://link.aps.org/doi/10.1103/PhysRevC.77.054319}
}

@article{PhysRevC.111.044606,
  title = {Low-energy upbend in the photon strength function of $^{57}\mathrm{Fe}$},
  author = {Kopeck\'y, J. and Tomandl, I.},
  journal = {Phys. Rev. C},
  volume = {111},
  issue = {4},
  pages = {044606},
  numpages = {5},
  year = {2025},
  month = {Apr},
  publisher = {American Physical Society},
  doi = {10.1103/PhysRevC.111.044606},
  url = {https://link.aps.org/doi/10.1103/PhysRevC.111.044606}
}

@article{PhysRevC.99.044308,
  title = {Constraints on the dipole photon strength functions from experimental multistep cascade spectra},
  author = {Krti\ifmmode \check{c}\else \v{c}\fi{}ka, M. and Goriely, S. and Hilaire, S. and P\'eru, S. and Valenta, S.},
  journal = {Phys. Rev. C},
  volume = {99},
  issue = {4},
  pages = {044308},
  numpages = {17},
  year = {2019},
  month = {Apr},
  publisher = {American Physical Society},
  doi = {10.1103/PhysRevC.99.044308},
  url = {https://link.aps.org/doi/10.1103/PhysRevC.99.044308}
}

@article{PhysRevLett.113.232502,
  title = {Novel technique for Constraining $r$-Process ($n$, $\ensuremath{\gamma}$) Reaction Rates},
  author = {Spyrou, A. and Liddick, S. N. and Larsen, A. C. and Guttormsen, M. and Cooper, K. and Dombos, A. C. and Morrissey, D. J. and Naqvi, F. and Perdikakis, G. and Quinn, S. J. and Renstr\o{}m, T. and Rodriguez, J. A. and Simon, A. and Sumithrarachchi, C. S. and Zegers, R. G. T.},
  journal = {Phys. Rev. Lett.},
  volume = {113},
  issue = {23},
  pages = {232502},
  numpages = {5},
  year = {2014},
  month = {Dec},
  publisher = {American Physical Society},
  doi = {10.1103/PhysRevLett.113.232502},
  url = {https://link.aps.org/doi/10.1103/PhysRevLett.113.232502}
}

@article{PhysRevLett.116.242502,
  title = {Experimental Neutron Capture Rate Constraint Far from Stability},
  author = {Liddick, S. N. and Spyrou, A. and Crider, B. P. and Naqvi, F. and Larsen, A. C. and Guttormsen, M. and Mumpower, M. and Surman, R. and Perdikakis, G. and Bleuel, D. L. and Couture, A. and Crespo Campo, L. and Dombos, A. C. and Lewis, R. and Mosby, S. and Nikas, S. and Prokop, C. J. and Renstrom, T. and Rubio, B. and Siem, S. and Quinn, S. J.},
  journal = {Phys. Rev. Lett.},
  volume = {116},
  issue = {24},
  pages = {242502},
  numpages = {6},
  year = {2016},
  month = {Jun},
  publisher = {American Physical Society},
  doi = {10.1103/PhysRevLett.116.242502},
  url = {https://link.aps.org/doi/10.1103/PhysRevLett.116.242502}
}

@article{PhysRevC.97.054329,
  title = {Enhanced low-energy $\ensuremath{\gamma}\text{-decay}$ strength of $^{70}\mathrm{Ni}$ and its robustness within the shell model},
  author = {Larsen, A. C. and Midtb\o{}, J. E. and Guttormsen, M. and Renstr\o{}m, T. and Liddick, S. N. and Spyrou, A. and Karampagia, S. and Brown, B. A. and Achakovskiy, O. and Kamerdzhiev, S. and Bleuel, D. L. and Couture, A. and Campo, L. Crespo and Crider, B. P. and Dombos, A. C. and Lewis, R. and Mosby, S. and Naqvi, F. and Perdikakis, G. and Prokop, C. J. and Quinn, S. J. and Siem, S.},
  journal = {Phys. Rev. C},
  volume = {97},
  issue = {5},
  pages = {054329},
  numpages = {9},
  year = {2018},
  month = {May},
  publisher = {American Physical Society},
  doi = {10.1103/PhysRevC.97.054329},
  url = {https://link.aps.org/doi/10.1103/PhysRevC.97.054329}
}

@article{SpyrouNatureComm2024,
  author  = {Spyrou, A. and Richman, D. and Couture, A. and Fields, C. E. and Liddick, S. N. and Childers, K. and Crider, B. P. and DeYoung, P. A. and Dombos, A. C. and Gastis, P. and Guttormsen, M. and Hermansen, K. and Larsen, A. C. and Lewis, R. and Lyons, S. and Midtb{\o}, J. E. and Mosby, S. and M{\"u}cher, D. and Naqvi, F. and Palmisano-Kyle, A. and Perdikakis, G. and Prokop, C. J. and Schatz, H. and Smith, M. K. and Sumithrarachchi, C. and Sweet, A.},
  title   = {Enhanced production of $^{60}$Fe in massive stars},
  journal = {Nat. Commun.},
  volume  = {15},
  pages   = {9608},
  year    = {2024},
  doi     = {10.1038/s41467-024-54040-4}
}

@article{PhysRevLett.119.052502,
  title = {Electric and Magnetic Dipole Strength at Low Energy},
  author = {Sieja, K.},
  journal = {Phys. Rev. Lett.},
  volume = {119},
  issue = {5},
  pages = {052502},
  numpages = {5},
  year = {2017},
  month = {Jul},
  publisher = {American Physical Society},
  doi = {10.1103/PhysRevLett.119.052502},
  url = {https://link.aps.org/doi/10.1103/PhysRevLett.119.052502}
}

@article{PhysRevC.100.024624,
  title = {Benchmarking the extraction of statistical neutron capture cross sections on short-lived nuclei for applications using the $\ensuremath{\beta}$-Oslo method},
  author = {Liddick, S. N. and Larsen, A. C. and Guttormsen, M. and Spyrou, A. and Crider, B. P. and Naqvi, F. and Midtb\o{}, J. E. and Bello Garrote, F. L. and Bleuel, D. L. and Crespo Campo, L. and Couture, A. and Dombos, A. C. and Giacoppo, F. and G\"orgen, A. and Hadynska-Klek, K. and Hagen, T. W. and Ingeberg, V. W. and Kheswa, B. V. and Lewis, R. and Mosby, S. and Perdikakis, G. and Prokop, C. J. and Quinn, S. J. and Renstr\o{}m, T. and Rose, S. J. and Sahin, E. and Siem, S. and Tveten, G. M. and Wiedeking, M. and Zeiser, F.},
  journal = {Phys. Rev. C},
  volume = {100},
  issue = {2},
  pages = {024624},
  numpages = {12},
  year = {2019},
  month = {Aug},
  publisher = {American Physical Society},
  doi = {10.1103/PhysRevC.100.024624},
  url = {https://link.aps.org/doi/10.1103/PhysRevC.100.024624}
}

@article{bnl_50v,
    title = {Nuclear Data Sheets for A=50},
    journal = {Nuclear Data Sheets},
    volume = {157},
    pages = {1-259},
    year = {2019},
    issn = {0090-3752},
    doi = {10.1016/j.nds.2019.04.001},
    author = {Chen, Jun and Singh, Balraj}
}

@book{suhonen,
  author    = {Suhonen, Jouni},
  title     = {From Nucleons to Nucleus},
  publisher = {Springer, Berlin, Heidelberg},
  year      = {2007}
}

@article{pygmy1,
  author  = {Bracco, A. and Crespi, F. C. L. and Lanza, E. G.},
  title   = {Gamma decay of pygmy states from inelastic scattering of ions},
  journal = {Eur. Phys. J. A},
  volume  = {51},
  number  = {8},
  pages   = {99},
  year    = {2015},
  doi     = {10.1140/epja/i2015-15099-6}
}

@article{pygmy2,
  author  = {Savran, D. and Aumann, T. and Zilges, A.},
  title   = {Experimental studies of the Pygmy Dipole Resonance},
  journal = {Prog. Part. Nucl. Phys.},
  volume  = {70},
  pages   = {210--245},
  year    = {2013},
  doi     = {10.1016/j.ppnp.2013.02.003}
}

@article{scissors,
  author  = {Schiller, A. and Voinov, A. and Algin, E. and Becker, J. A. and Bernstein, L. A. and Garrett, P. E. and Guttormsen, M. and Nelson, R. O. and Rekstad, J. and Siem, S.},
  title   = {Low-energy $M1$ excitation mode in $^{172}$Yb},
  journal = {Phys. Lett. B},
  volume  = {633},
  number  = {2},
  pages   = {225--230},
  year    = {2006},
  doi     = {10.1016/j.physletb.2005.12.043}
}

@article{PhysRevC.76.044303,
  title = {Nuclear level densities and \ensuremath{\gamma}-ray strength functions in $^{44,45}\mathrm{Sc}$},
  author = {Larsen, A. C. and Guttormsen, M. and Chankova, R. and Ingebretsen, F. and L\"onnroth, T. and Messelt, S. and Rekstad, J. and Schiller, A. and Siem, S. and Syed, N. U. H. and Voinov, A.},
  journal = {Phys. Rev. C},
  volume = {76},
  issue = {4},
  pages = {044303},
  numpages = {11},
  year = {2007},
  month = {Oct},
  publisher = {American Physical Society},
  doi = {10.1103/PhysRevC.76.044303},
  url = {https://link.aps.org/doi/10.1103/PhysRevC.76.044303}
}

@article{PhysRevC.98.064321,
  title = {Consolidating the concept of low-energy magnetic dipole decay radiation},
  author = {Midtb\o{}, J. E. and Larsen, A. C. and Renstr\o{}m, T. and Bello Garrote, F. L. and Lima, E.},
  journal = {Phys. Rev. C},
  volume = {98},
  issue = {6},
  pages = {064321},
  numpages = {12},
  year = {2018},
  month = {Dec},
  publisher = {American Physical Society},
  doi = {10.1103/PhysRevC.98.064321},
  url = {https://link.aps.org/doi/10.1103/PhysRevC.98.064321}
}

@article{PhysRevC.41.1941,
  title = {Test of gamma-ray strength functions in nuclear reaction model calculations},
  author = {Kopecky, J. and Uhl, M.},
  journal = {Phys. Rev. C},
  volume = {41},
  issue = {5},
  pages = {1941--1955},
  numpages = {0},
  year = {1990},
  month = {May},
  publisher = {American Physical Society},
  doi = {10.1103/PhysRevC.41.1941},
  url = {https://link.aps.org/doi/10.1103/PhysRevC.41.1941}
}

@book{lawson1980theory,
  author    = {Lawson, R. D.},
  title     = {Theory of the Nuclear Shell Model},
  publisher = {Clarendon Press, Oxford},
  year      = {1980}
}

@book{rin80,
  author    = {Ring, P. and Schuck, P.},
  title     = {The Nuclear Many-Body Problem},
  publisher = {Springer-Verlag, New York},
  year      = {1980}
}

@incollection{Bartholomew1973,
  author    = {Bartholomew, G. A. and Earle, E. D. and Ferguson, A. J. and Knowles, J. W. and Lone, M. A.},
  title     = {Gamma-Ray Strength Functions},
  booktitle = {Advances in Nuclear Physics, Vol. 7},
  editor    = {Baranger, Michel and Vogt, Erich},
  publisher = {Springer, Boston, MA},
  pages     = {229--324},
  year      = {1973},
  doi       = {10.1007/978-1-4615-9044-6_4}
}

@article{GLOECKNER1974313,
  author  = {Gloeckner, D. H. and Lawson, R. D.},
  title   = {Spurious center-of-mass motion},
  journal = {Phys. Lett. B},
  volume  = {53},
  number  = {4},
  pages   = {313--318},
  year    = {1974},
  doi     = {10.1016/0370-2693(74)90390-6}
}

@article{PhysRevC.95.045805,
  title = {$^{137,138,139}\mathbf{La}(n,\ensuremath{\gamma})$ cross sections constrained with statistical decay properties of $^{138,139,140}\mathbf{La}$ nuclei},
  author = {Kheswa, B. V. and Wiedeking, M. and Brown, J. A. and Larsen, A. C. and Goriely, S. and Guttormsen, M. and Bello Garrote, F. L. and Bernstein, L. A. and Bleuel, D. L. and Eriksen, T. K. and Giacoppo, F. and G\"orgen, A. and Goldblum, B. L. and Hagen, T. W. and Koehler, P. E. and Klintefjord, M. and Malatji, K. L. and Midtb\o{}, J. E. and Nyhus, H. T. and Papka, P. and Renstr\o{}m, T. and Rose, S. J. and Sahin, E. and Siem, S. and Tornyi, T. G.},
  journal = {Phys. Rev. C},
  volume = {95},
  issue = {4},
  pages = {045805},
  numpages = {9},
  year = {2017},
  month = {Apr},
  publisher = {American Physical Society},
  doi = {10.1103/PhysRevC.95.045805},
  url = {https://link.aps.org/doi/10.1103/PhysRevC.95.045805}
}

@article{BROWN2014115,
  author  = {Brown, B. A. and Rae, W. D. M.},
  title   = {The shell-model code NuShellX@MSU},
  journal = {Nucl. Data Sheets},
  volume  = {120},
  pages   = {115--118},
  year    = {2014},
  doi     = {10.1016/j.nds.2014.07.022}
}

@techreport{OXBASH,
  author      = {Brown, B. A. and Etchegoyen, A. and Rae, W. D. M. and Godwin, N. S. and Richter, W. A. and Zimmerman, C. H. and Ormand, W. E. and Winfield, J. S.},
  title       = {OXBASH for Windows},
  institution = {National Superconducting Cyclotron Laboratory, Michigan State University},
  number      = {524},
  year        = {1985}
}

@article{ERICSON1959,
  author  = {Ericson, Torleif},
  title   = {A statistical analysis of excited nuclear states},
  journal = {Nucl. Phys.},
  volume  = {11},
  pages   = {481--491},
  year    = {1959},
  doi     = {10.1016/0029-5582(59)90291-3}
}

@article{Brown2014,
  title = {Large Low-Energy $M1$ Strength for $^{56,57}\mathrm{Fe}$ within the Nuclear Shell Model},
  author = {Brown, B. Alex and Larsen, A. C.},
  journal = {Phys. Rev. Lett.},
  volume = {113},
  issue = {25},
  pages = {252502},
  numpages = {5},
  year = {2014},
  month = {Dec},
  publisher = {American Physical Society},
  doi = {10.1103/PhysRevLett.113.252502},
  url = {https://link.aps.org/doi/10.1103/PhysRevLett.113.252502}
}

@article{PhysRevC.69.034335,
  title = {New effective interaction for $pf$-shell nuclei and its implications for the stability of the $N=Z=28$ closed core},
  author = {Honma, M. and Otsuka, T. and Brown, B. A. and Mizusaki, T.},
  journal = {Phys. Rev. C},
  volume = {69},
  issue = {3},
  pages = {034335},
  numpages = {34},
  year = {2004},
  month = {Mar},
  publisher = {American Physical Society},
  doi = {10.1103/PhysRevC.69.034335},
  url = {https://link.aps.org/doi/10.1103/PhysRevC.69.034335}
}

@article{PhysRevC.95.024322,
  title = {Low energy magnetic radiation enhancement in the ${f}_{7/2}$ shell},
  author = {Karampagia, S. and Brown, B. A. and Zelevinsky, V.},
  journal = {Phys. Rev. C},
  volume = {95},
  issue = {2},
  pages = {024322},
  numpages = {7},
  year = {2017},
  month = {Feb},
  publisher = {American Physical Society},
  doi = {10.1103/PhysRevC.95.024322},
  url = {https://link.aps.org/doi/10.1103/PhysRevC.95.024322}
}

@article{PhysRevC.97.054321,
  title = {Systematic shell-model study of $\ensuremath{\beta}$-decay properties and Gamow-Teller strength distributions in $A\ensuremath{\approx}40$ neutron-rich nuclei},
  author = {Yoshida, Sota and Utsuno, Yutaka and Shimizu, Noritaka and Otsuka, Takaharu},
  journal = {Phys. Rev. C},
  volume = {97},
  issue = {5},
  pages = {054321},
  numpages = {17},
  year = {2018},
  month = {May},
  publisher = {American Physical Society},
  doi = {10.1103/PhysRevC.97.054321},
  url = {https://link.aps.org/doi/10.1103/PhysRevC.97.054321}
}

@article{Rauscher_2002,
doi = {10.1086/341728},
url = {https://dx.doi.org/10.1086/341728},
year = {2002},
month = {sep},
publisher = {},
volume = {576},
number = {1},
pages = {323},
author = {Rauscher, T. and Heger, A. and Hoffman, R. D. and Woosley, S. E.},
title = {Nucleosynthesis in Massive Stars with Improved Nuclear and Stellar Physics},
journal = {The Astrophysical Journal}
}

@article{Imasheva_2023,
    author = {Imasheva, Liliya and Janka, Hans-Thomas and Weiss, Achim},
    title = {Parametrizations of thermal bomb explosions for core-collapse supernovae and 56Ni production},
    journal = {Monthly Notices of the Royal Astronomical Society},
    volume = {518},
    number = {2},
    pages = {1818-1839},
    year = {2022},
    month = {11},
    issn = {0035-8711},
    doi = {10.1093/mnras/stac3239},
    url = {https://doi.org/10.1093/mnras/stac3239},
    eprint = {https://academic.oup.com/mnras/article-pdf/518/2/1818/47224503/stac3239.pdf},
}

@article{DJALALI19821,
title = {Systematics of the excitation of M1 resonances in medium heavy nuclei by 200 MeV proton inelastic scattering},
journal = {Nuclear Physics A},
volume = {388},
number = {1},
pages = {1-18},
year = {1982},
issn = {0375-9474},
doi = {https://doi.org/10.1016/0375-9474(82)90505-X},
url = {https://www.sciencedirect.com/science/article/pii/037594748290505X},
author = {C. Djalali and N. Marty and M. Morlet and A. Willis and J.C. Jourdain and N. Anantaraman and G.M. Crawley and A. Galonsky and P. Kitching}
}

@article{POVES2001157,
title = {Shell model study of the isobaric chains A=50, A=51 and A=52},
journal = {Nuclear Physics A},
volume = {694},
number = {1},
pages = {157-198},
year = {2001},
issn = {0375-9474},
doi = {https://doi.org/10.1016/S0375-9474(01)00967-8},
url = {https://www.sciencedirect.com/science/article/pii/S0375947401009678},
author = {A. Poves and J. Sánchez-Solano and E. Caurier and F. Nowacki}
}

@article{usd,
  title = {},
  author = {Brown, B.A. and Wildenthal, B.H.},
  journal = {Annu. Rev. Nucl. Part. Sci.},
  volume = {38},
  issue = {},
  pages = {29},
  numpages = {},
  year = {1988},
  month = {},
  publisher = {},
  doi = {},
  url = {}
}

@article{gxpf1b,
 title = {Shell-model description of neutron-rich Ca isotopes},
 author = {Honma, M. and Otsuka, T. and Mizusaki, T.},
 journal = {RIKEN Accel. Prog. Rep.},
 volume = {41},
 pages = {32},
 year = {2008},
}

@article{vmu,   
  title = {Novel Features of Nuclear Forces and Shell Evolution in Exotic Nuclei},
  author = {Otsuka, Takaharu and Suzuki, Toshio and Honma, Michio and Utsuno, Yutaka and Tsunoda, Naofumi and Tsukiyama, Koshiroh and Hjorth-Jensen, Morten},
  journal = {Phys. Rev. Lett.},
  volume = {104},
  issue = {1},
  pages = {012501},
  numpages = {4},
  year = {2010},
  month = {Jan},
  publisher = {American Physical Society},
  doi = {10.1103/PhysRevLett.104.012501},
  url = {http://link.aps.org/doi/10.1103/PhysRevLett.104.012501}
}

@article{gxpf1,
  title = {New effective interaction for $pf$-shell nuclei and its implications for the stability of the $N=Z=28$ closed core},
  author = {Honma, M. and Otsuka, T. and Brown, B. A. and Mizusaki, T.},
  journal = {Phys. Rev. C},
  volume = {69},
  issue = {3},
  pages = {034335},
  numpages = {34},
  year = {2004},
  month = {Mar},
  publisher = {American Physical Society},
  doi = {10.1103/PhysRevC.69.034335},
  url = {https://link.aps.org/doi/10.1103/PhysRevC.69.034335}
}

@misc{DahlZenodo2026,
  author       = {Dahl, J. K.},
  title        = {Supplemental Material: Microscopic Study of the Low-Energy Enhancement in the Gamma-Decay Strength of $^{50}$V},
  howpublished = {Zenodo},
  year         = {2026},
  doi          = {10.5281/zenodo.18403638}
}
\bibliographystyle{apsrev4-2}
\end{document}